\documentclass[letterpaper,aps,prd,twocolumn,tightenlines,preprintnumbers,nofootinbib,showkeys,superscriptaddress,longbibliography]{revtex4-1}
\pdfoutput=1

\usepackage{XCharter}
\usepackage{mathptmx}
\usepackage[T1]{fontenc}
\usepackage{blkarray}
\usepackage{fullpage}
\usepackage{amsfonts}
\usepackage{amsmath}
\usepackage{slashed}
\usepackage{amssymb}
\usepackage{graphicx}
\usepackage{epic}
\usepackage{eepic}
\usepackage{epsfig}
\usepackage{latexsym}
\usepackage[dvipsnames]{xcolor}
\usepackage[export]{adjustbox}
\usepackage{float}
\usepackage{multirow}
\usepackage{hyperref}
\usepackage{enumitem}
\usepackage{lipsum}
\hypersetup{colorlinks=true,citecolor=red,linkcolor=NavyBlue,urlcolor=NavyBlue}
\usepackage[caption=false,labelformat=simple]{subfig}
 
\usepackage{natbib}
\usepackage{relsize}
\usepackage[left=1.6cm,right=1.6cm,top=1.65cm,bottom=1.65cm]{geometry}
\usepackage{comment}
\usepackage{shorthand}
\usepackage{dcolumn}
\usepackage{longtable}

\usepackage{pifont}
\begin{document}

\title{Fresh look at the LHC limits on scalar leptoquarks}

\author{Arvind Bhaskar}
\email{arvind.bhaskar@iopb.res.in}
\affiliation{Center for Computational Natural Sciences and Bioinformatics, International Institute of Information Technology, Hyderabad 500 032, India}
\affiliation{Institute of Physics, Sachivalaya Marg, Bhubaneswar 751005, India}

\author{Arijit Das}
\email{arijit21@iisertvm.ac.in}
\affiliation{Indian Institute of Science Education and Research Thiruvananthapuram, Vithura, Kerala, 695 551, India}

\author{Tanumoy Mandal}
\email{tanumoy@iisertvm.ac.in}
\affiliation{Indian Institute of Science Education and Research Thiruvananthapuram, Vithura, Kerala, 695 551, India}

\author{Subhadip Mitra}
\email{subhadip.mitra@iiit.ac.in}
\affiliation{Center for Computational Natural Sciences and Bioinformatics, International Institute of Information Technology, Hyderabad 500 032, India}
\affiliation{Center for Quantum Science and Technology, International Institute of Information Technology, Hyderabad 500 032, India}

\author{Rachit Sharma}
\email{rachit21@iisertvm.ac.in}
\affiliation{Indian Institute of Science Education and Research Thiruvananthapuram, Vithura, Kerala, 695 551, India}

\begin{abstract}
\noindent
The scalar-leptoquark (sLQ) parameter space is well explored experimentally. The direct pair production searches at the LHC have excluded light sLQs almost model agnostically, and the high-$p_{\rm T}$ dilepton tail data have put strong bounds on the leptoquark-quark-lepton Yukawa couplings for a wide range of sLQ masses. However, these do not show the complete picture. Previously,  Mandal \emph{et al.} [Single productions of colored particles at the LHC: An example with scalar leptoquarks, \href{https://doi.org/10.1007/JHEP07(2015)028}{J. High Energy Phys. 07 (2015) 028}] showed how the dilepton-dijet data from the pair production searches could give strong limits on these couplings. This was possible by including the single-production contribution to the dilepton-dijet signal. In this paper, we take a fresh look at the LHC limits on all sLQs by following the same principle and combine all significant contributions---from pair and single productions, $t$-channel sLQ exchange and its interference with the Standard Model background---to the $\mu\mu jj$ final state and recast the limits. We notice that the sLQ exchange and its interference with the background processes play significant roles in the limits. The $\mu\mu jj$-recast  limits are comparable to or, in some cases, significantly better than the currently known limits (from high-$p_{\rm T}$ dilepton data and direct searches), i.e., the LHC data rules out more parameter space than what is considered in the current literature. For the first time, we also show how including the QED processes can noticeably improve the sLQ mass exclusion limits from the QCD-only limits.

\end{abstract}

\maketitle 

%%%%%%%%%%%%%%%%%%%%%%%%%%%%%%%%%%%%%%%%%%%%%%%%%%
\section{Introduction}\label{sec:intro}
\noindent
Hypothetical bosons called leptoquarks (LQs) have found significant importance in theoretical and experimental particle physics. These scalar/vector colour-triplet particles carry fractional electric charges and connect the Standard Model (SM) quark and lepton sectors. Many beyond-the-SM (BSM) theories predict their existence at the TeV scale~\cite{Pati:1974yy,Georgi:1974sy,Fritzsch:1974nn,Farhi:1980xs,Schrempp:1984nj,Wudka:1985ef,Barbier:2004ez}. Their current popularity in the literature mainly stems from their roles in %simultaneously 
resolving the flavour and other persistent anomalies (for example, muon $g-2$~\cite{Muong-2:2021vma,Muong-2:2021ojo,Muong-2:2023cdq}). This has motivated phenomenologists to speculate on new top-down/bottom-up models~\cite{Goncalves:2023qpz,DaRold:2023hmx} and propose novel collider signatures and search strategies for LQs~\cite{Blumlein:1996qp,Dorsner:2014axa,Diaz:2017lit,Bandyopadhyay:2018syt,Schmaltz:2018nls,Bhaskar:2020kdr,Greljo:2020tgv,Dorsner:2021chv,Bandyopadhyay:2021pld,Bhaskar:2022ygp,Parashar:2022wrd,Ghosh:2023ocz,Florez:2023jdb,Arganda:2023qni}. The strong theoretical motivation for the low-mass LQ states within the reach of the Large Hadron Collider (LHC) makes their search an active and popular pursuit among experimental particle physicists. In recent years, research in these directions has become progressively comprehensive. The ATLAS and CMS collaborations have conducted extensive searches for the pair and single productions of LQs (see the \href{https://cds.cern.ch/record/2852994/files/ATL-PHYS-PUB-2023-006.pdf}{ATLAS} and \href{https://twiki.cern.ch/twiki/pub/CMSPublic/SummaryPlotsEXO13TeV/barplot_EQ_MUQ_TAUQ_NUQ_v5.pdf}{CMS} summary plots for LQ searches and the references therein).  In the absence of any discovery, these searches have put strong limits on LQ parameters. 

Most commonly, these searches focus on LQ pair production (PP), where the LQs decay into lepton-jet pairs forming dilepton-dijet ($\ell\ell jj$, where $\ell=e,\mu$) final states. From these, model-independent limits on the LQ mass are drawn assuming the LQ-quark-lepton Yukawa couplings ($y$) responsible for LQ decays are small and the PP process is essentially strong-interaction mediated in this limit. The current upper limit on the PP cross section in the $\ell\ell jj$ channel is about $0.1$~fb for scalar LQ (sLQ) masses around $1.5$~TeV~\cite{ATLAS:2020dsk}. The cross sections of single production (SP) processes depend on the new coupling(s), and hence direct SP searches in the $\ell\ell j$ final states provide LQ mass-dependent upper limits on $y$ (which can also be interpreted as $y$-dependent lower bounds on LQ masses).

Apart from the direct searches, there are also some less obvious sources of LHC bounds. For example, the high-$p_{\rm T}$ tail of the dilepton or the lepton + $\slashed E_{\rm T}$ resonance (like $Z^\prime$ or $W^\prime$) search data can limit the LQ parameter space~\cite{Raj:2016aky,Greljo:2017vvb,Bansal:2018eha,Alvarez:2018jfb,Alves:2018krf,Fuentes-Martin:2020lea,Buonocore:2020erb,Crivellin:2021egp,Haisch:2022lkt,Allwicher:2022gkm}. In the quark fusion processes, a $t$-channel LQ exchange can produce two high-$p_{\rm T}$ leptons (or a lepton and a neutrino) affecting the tails of the distributions. The $t$-channel LQ exchange [we refer to it as the indirect production (IP)] is highly sensitive to the new couplings as the amplitude is proportional to the second power in the Yukawa couplings. Hence, one can use the high-$p_{\rm T}$ dilepton or lepton + $\slashed E_{\rm T}$ data to constrain $y$ for a range of masses of the LQ involved (wider than the mass range accessible to the direct searches). Interestingly, the IP amplitude interferes with the gauge bosons-mediated dilepton or monolepton + $\slashed E_{\rm T}$ productions; the interference terms play major roles in determining the limits on the couplings~\cite{Mandal:2018kau}.

In principle, we can combine all the latest ATLAS and CMS bounds to obtain the allowed regions in the LQ parameter space. That, however, will only show a partial picture for a couple of reasons. Just as we recast the dilepton data to draw limits on the LQ parameters, we can recast the direct-search data by including the contribution of all processes in the signal to draw stronger limits. If some of the new Yukawa couplings are $\mc{O}(1)$ (as needed by the anomalies), the $\ell\ell j j$ signal can receive significant contributions from the SP and IP (and its interference) processes and lead to stronger (albeit model-dependent) bounds than the model-independent PP bounds. This is not entirely a new observation---we illustrated how systematic inclusion of SP events in the PP-search ($\ell\ell jj$) signal and PP events in the SP-search ($\ell\ell j$) signal lead to stronger limits in Ref.~\cite{Mandal:2015vfa} (also see Refs.~\cite{Aydemir:2019ynb,Chandak:2019iwj,Bhaskar:2021pml,Bhaskar:2021gsy,Bhaskar:2022vgk,Aydemir:2022lrq}). However, this point---not specific to LQs but applicable to many other BSM searches as well~\cite{Mandal:2012rx,Mandal:2016csb}---has largely been overlooked in the literature. Second, these limits ignore the contributions of a class of diagrams. Since LQs have both electric and colour charges, they can couple to both photons and gluons. Interestingly, the gauge symmetries allow for a gluon-photon-sLQ-sLQ interaction~\cite{Crivellin:2021ejk}. So far, the contributions from the photon-initiated diagrams (or those involving this particular interaction term) are not accounted for in the collider analyses in general. 

Given the important roles the TeV-scale LQs play in a wide range of BSM scenarios addressing various open problems, it is crucial to scrutinise the LHC limits on them in detail. These limits are independent of (and competitive to) other low-energy bounds (like those from various flavour observables)---they exclude various LQ scenarios which are allowed by the low-energy limits (see, e.g., Ref.~\cite{Bhaskar:2021pml}). In this paper, we reexamine the LHC limits on the scalar LQs from the latest ($\sim 140$ fb$^{-1}$) data in light of the above points. Since, at the LHC, muons are the easiest leptons to identify, we consider the $\m\m jj$~\cite{ATLAS:2020dsk} and $\m\m$~\cite{CMS:2021ctt} data for our purpose (there is no $\m\m j$ data at this luminosity). In addition, we also consider the latest $\mu b \mu b$~\cite{ATLAS:2020dsk} and $\mu t \mu t$~\cite{CMS:2022nty} data in recasting the exclusion limits.

The plan for the rest of the paper is as follows. We review the interactions and different production mechanisms for the sLQs that can couple with the muon in Sec.~\ref{sec:slqprod}; explain the recast methodologies in Sec.~\ref{sec:recast}; present the recast limits in Sec.~\ref{sec:el}; and finally, we conclude in Sec.~\ref{sec:conclu}.

%%%%%%%%%%%%%%%%%%%%%%%%%%%%%%%%%%%%%%%%%%%%%%%%%%%%%%%%%%%%%%%%%%
\begin{table*}
\caption{The Yukawa interactions in up and down-aligned scalar-LQ scenarios (in the notation of Ref.~\cite{Dorsner:2016wpm}, except for the $R_2$ couplings where the roles of $i$ and $j$ are interchanged). We ignore the diquark interactions as those are not relevant to our analysis. Here, $V\equiv V_{\rm CKM}$ denotes the Cabibbo-Kobayashi-Maskawa matrix. In the rest of the paper, we use a slightly different but more explicit notation for the Yukawa couplings---see Appendix~\ref{appendix:notation} for details. \label{tab:SLQYuk}  }
\centering{\footnotesize\renewcommand\baselinestretch{2}\selectfont
\begin{tabular*}{\textwidth}{l @{\extracolsep{\fill}}cc}
\hline Model & Down-aligned Yukawa Interactions & Up-aligned Yukawa Interactions \\ \hline\hline
$S_{1}$     & $-y^{LL}_{1ij}\ \overline{d^C_L}^{i} \nu^{j}_{L}S_{1} + (V^{*}y_{1}^{LL})_{ij}\ \overline{u^C_L}^{i} e^{j}_{L}S_{1} + y^{RR}_{1ij}\ \overline{u^C_R}^{i}e^{j}_{R} S_{1}$            & 
$-(V^{T}y^{LL}_1)_{ij}\ \overline{d^C_L}^{i} \nu^{j}_{L}S_{1} + y^{LL}_{1ij}\ \overline{u^C_L}^{i} e^{j}_{L}S_{1} + y^{RR}_{1ij}\ \overline{u^C_R}^{i} e^{j}_{R}S_{1}$ \\ 
$\widetilde{S}_{1}$     & \multicolumn{2}{c}{$\widetilde{y}^{RR}_{1ij}\ \overline{d^C_{R}}^{i}e^{j}_{R}\widetilde{S}_{1}$} \\ 
\multirow{2}{*}{$R_{2}$}     & $-y^{RL}_{2ij}\ (\overline{u}_{R}^{i}e^{j}_{L}R^{5/3}_{2} - \overline{u}_{R}^{i}\nu^{j}_{L}R^{2/3}_{2})$            & $-y^{RL}_{2ij}\ (\overline{u}_{R}^{i}e^{j}_{L}R^{5/3}_{2} - \overline{u}_{R}^{i}\nu^{j}_{L}R^{2/3}_{2})$          \\ 
&$ + (Vy_{2}^{LR})_{ij}\ \overline{u}_L^{i}e^{j}_{R}R^{5/3}_{2} + y^{{LR}}_{2ij}\ \overline{d}_L^{i}e^{j}_{R}R^{2/3}_{2}$&$ + y^{LR}_{2ij}\ \overline{u}_L^{i}e^{j}_{R}R^{5/3}_{2} + (V^{\dagger}y_{2}^{LR})_{ij}\ \overline{d}_L^{i}e^{j}_{R}R^{2/3}_{2}$\\
$\widetilde{R}_{2}$     & \multicolumn{2}{c}{$-\widetilde{y}^{RL}_{2ij}\ 
(\overline{d}^{i}_{R}e^{j}_{L}\widetilde{R}^{2/3}_{2} - 
\overline{d}^{i}_{R}\nu^{j}_{L}\widetilde{R}^{-1/3}_{2})$} \\ 
 \multirow{2}{*}{$S_{3}$}     & $-y^{LL}_{3ij}\ \overline{d^{C}_{L}}^{i}\nu^{j}_{L}S^{1/3}_{3}-(V^{*}y_{3}^{LL})_{ij}\ \overline{u^{C}_{L}}^{i}e^{j}_{L}S^{1/3}_{3}$             & $-(V^Ty_{3}^{LL})_{ij}\ \overline{d^{C}_{L}}^{i}\nu^{j}_{L}S^{1/3}_{3}-y^{LL}_{3ij}\ \overline{u^{C}_{L}}^{i}e^{j}_{L}S^{1/3}_{3}$         \\ 
 &  $-\sqrt{2}y^{LL}_{3ij}\ \overline{d^{C}_{L}}^{i}e^{j}_{L}S^{4/3}_{3}+\sqrt{2}(V^* y^{LL}_{3})_{ij}\ \overline{u^{C}_{L}}^{i}\nu^{j}_{L}S^{-2/3}_{3}$ & $-\sqrt{2}(V^T y_{3}^{LL})_{ij}\ \overline{d^{C}_{L}}^{i}e^{j}_{L}S^{4/3}_{3}+\sqrt{2}y^{LL}_{3ij}\  \overline{u^{C}_{L}}^{i}\nu^{j}_{L}S^{-2/3}_{3}$ \\
\hline
\end{tabular*}}
\end{table*}
%%%%%%%%%%%%%%%%%%%%%%%%%%%%%%%%%%%%%%%%%%%%%%%%%%%%%%%%%%%%%%%%%%
%%%%%%%%%%%%%%%%%%%%%%%%%%%%%%%%%%%%%%%%%%%%%%%%%%%%%%%%%%%%%%%%%%
\begin{figure*}
\centering
\captionsetup[subfigure]{labelformat=empty}
\subfloat[(a)]{\includegraphics[width=0.19\textwidth]{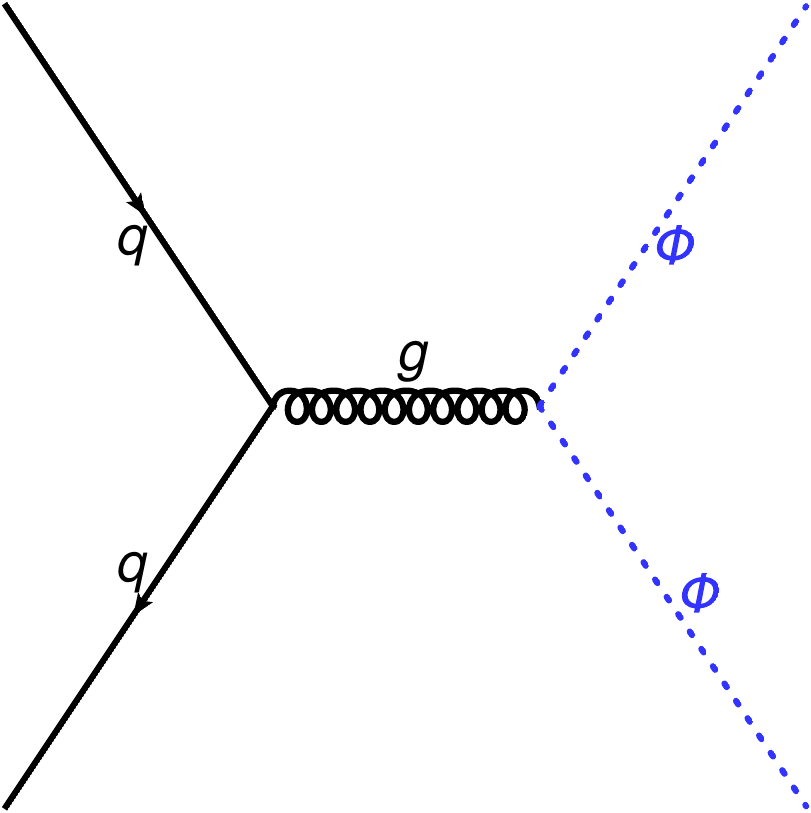}\label{qqlqlq}}\hfill
\subfloat[(b)]{\includegraphics[width=0.19\textwidth]{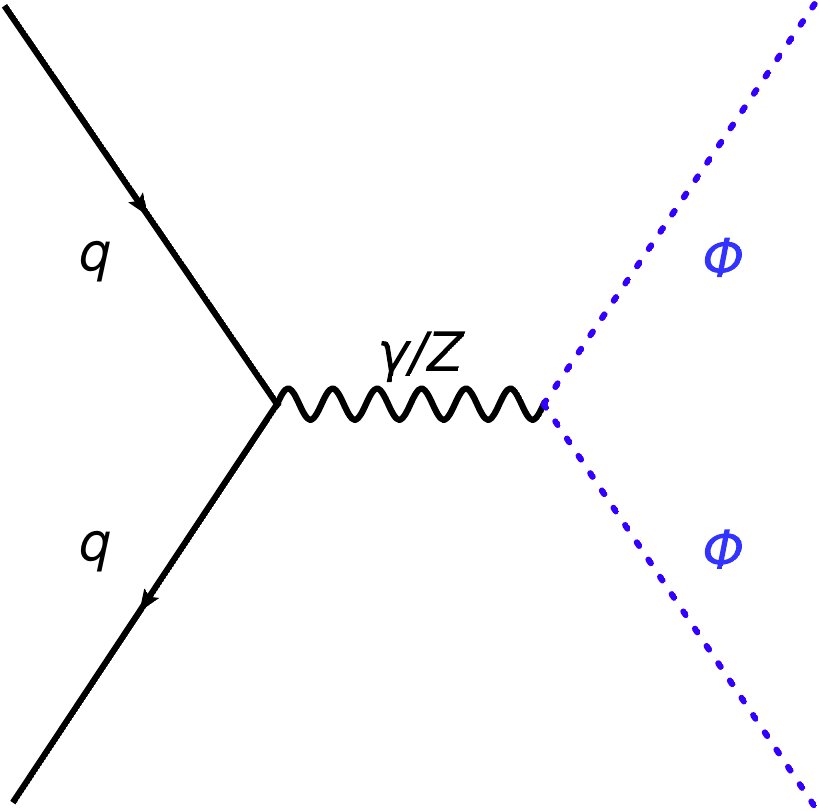}\label{qqzlqlq}}\hfill
\subfloat[(c)]{\includegraphics[width=0.19\textwidth]{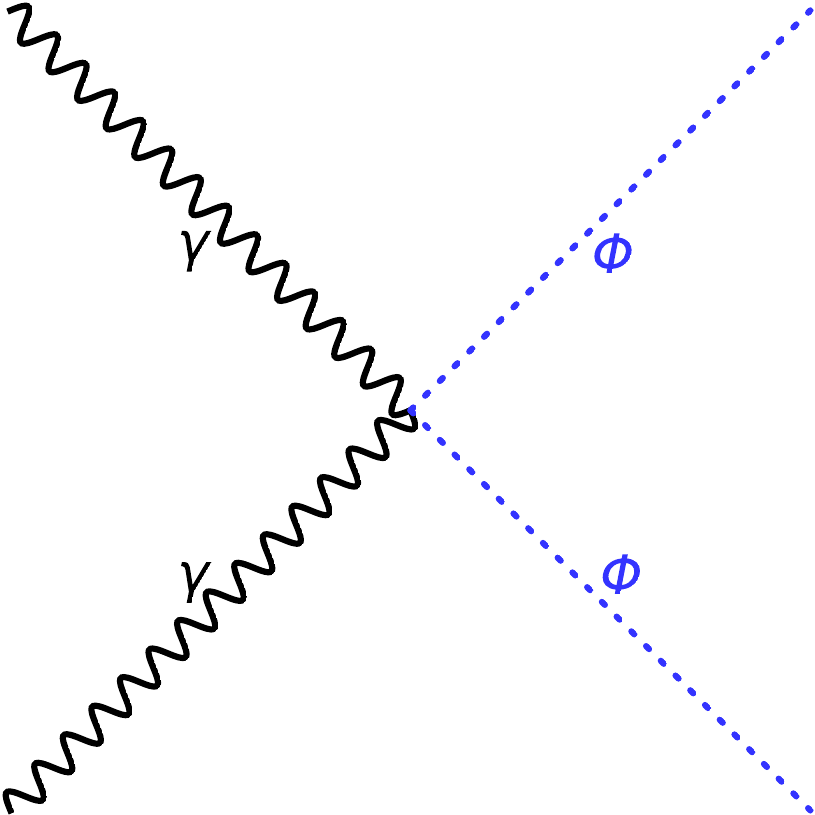}\label{pplqlq}}\hfill
\subfloat[(d)]{\includegraphics[width=0.19\textwidth]{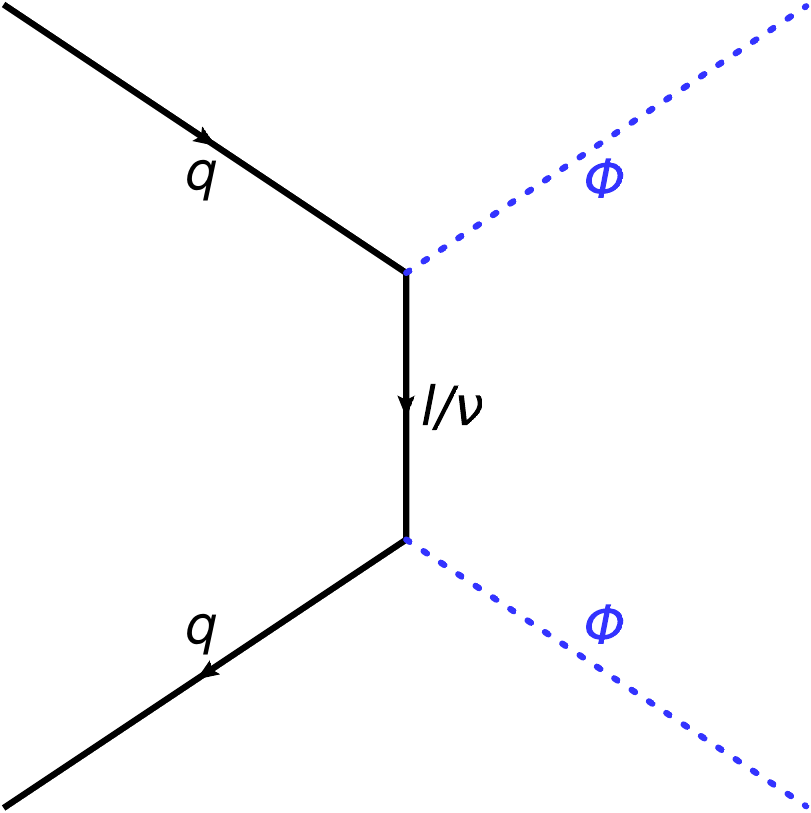}\label{qqtch}}
\\
\subfloat[(e)]{\includegraphics[width=0.19\textwidth]{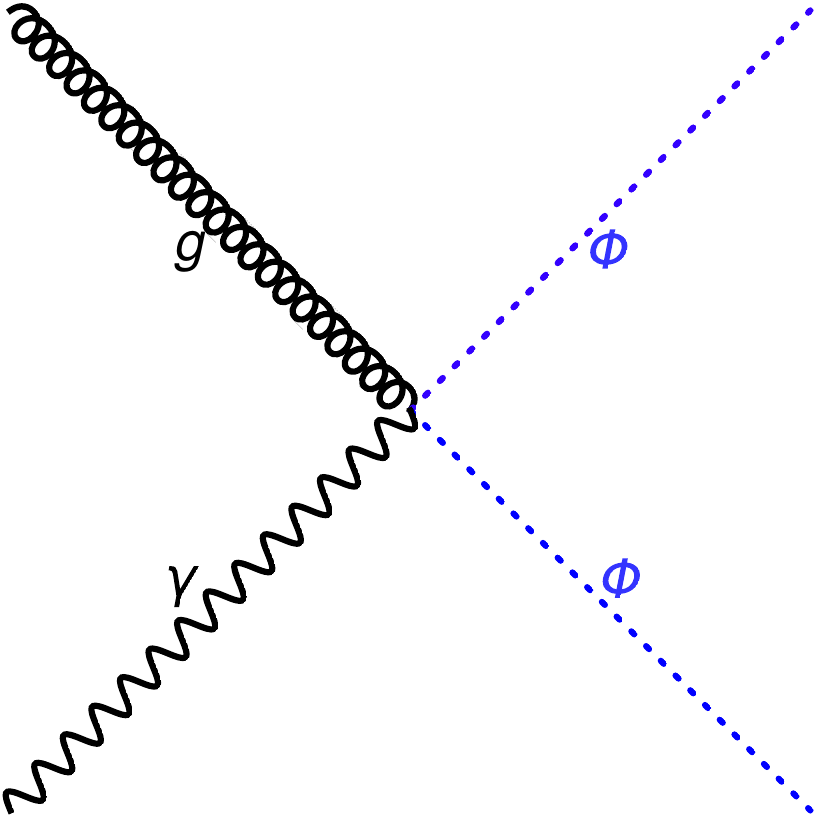}\label{gglqlq3pt}}\hfill
\subfloat[(f)]{\includegraphics[width=0.19\textwidth]{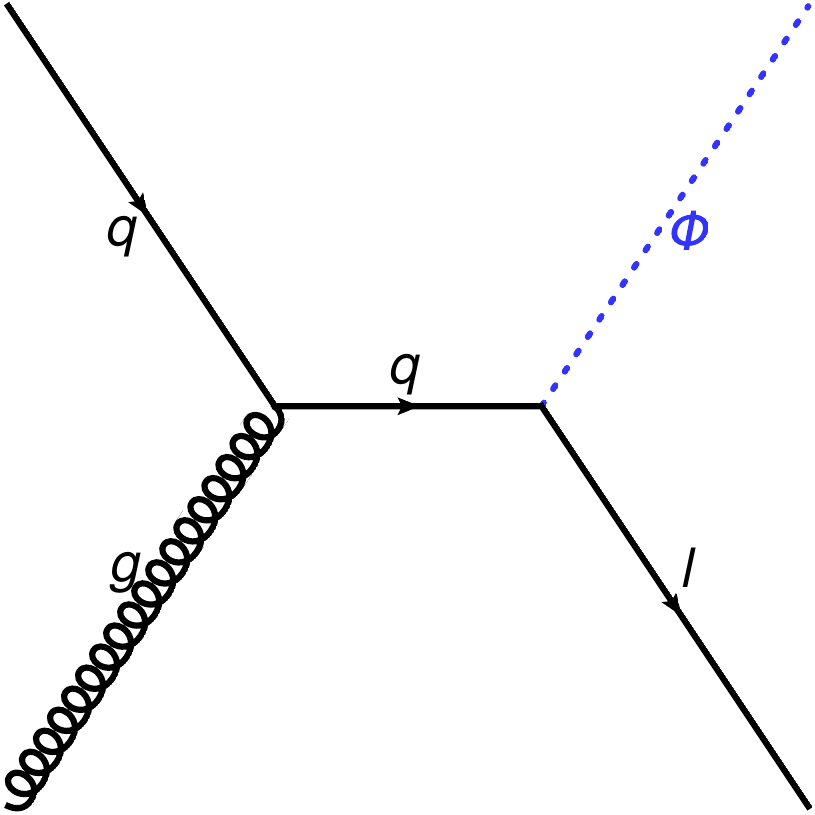}\label{2bspqg}}\hfill
\subfloat[(g)]{\includegraphics[width=0.19\textwidth]{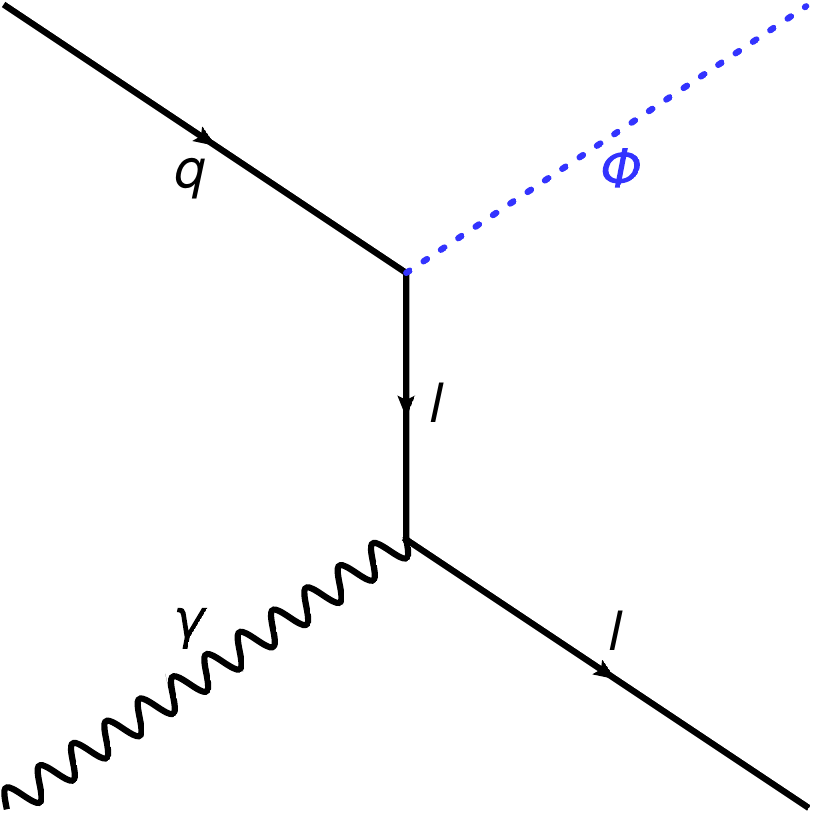}\label{2bspqgamma}}\hfill
\subfloat[(h)]{\includegraphics[width=0.19\textwidth]{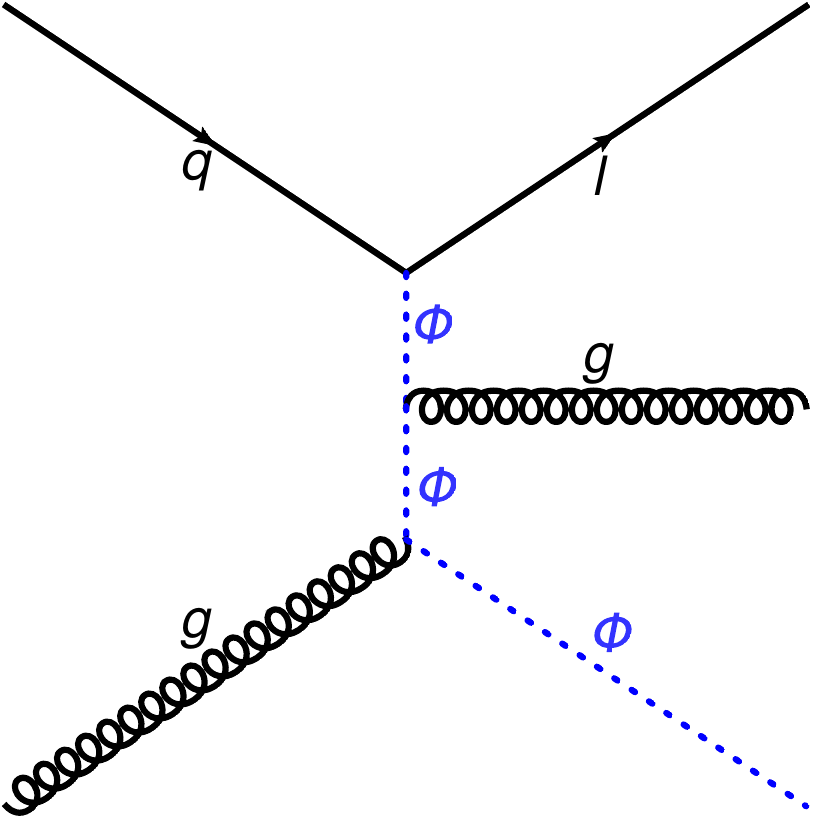}\label{SPqg}}
\\
\subfloat[(i)]{\includegraphics[width=0.19\textwidth]{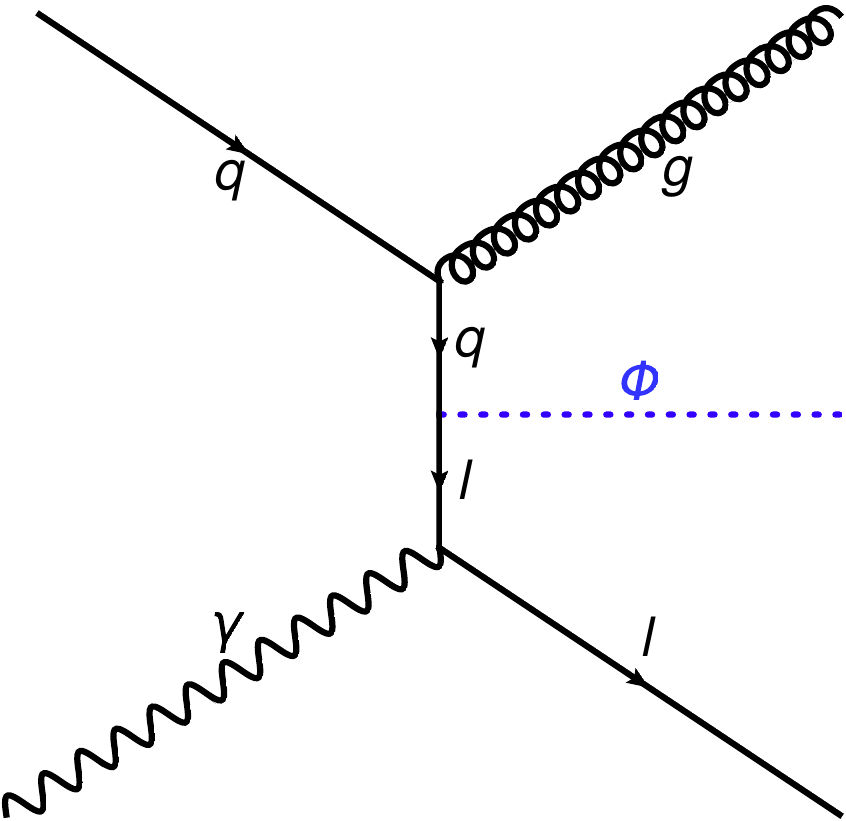}\label{SPqp}}\hfill
\subfloat[(j)]{\includegraphics[width=0.19\textwidth]{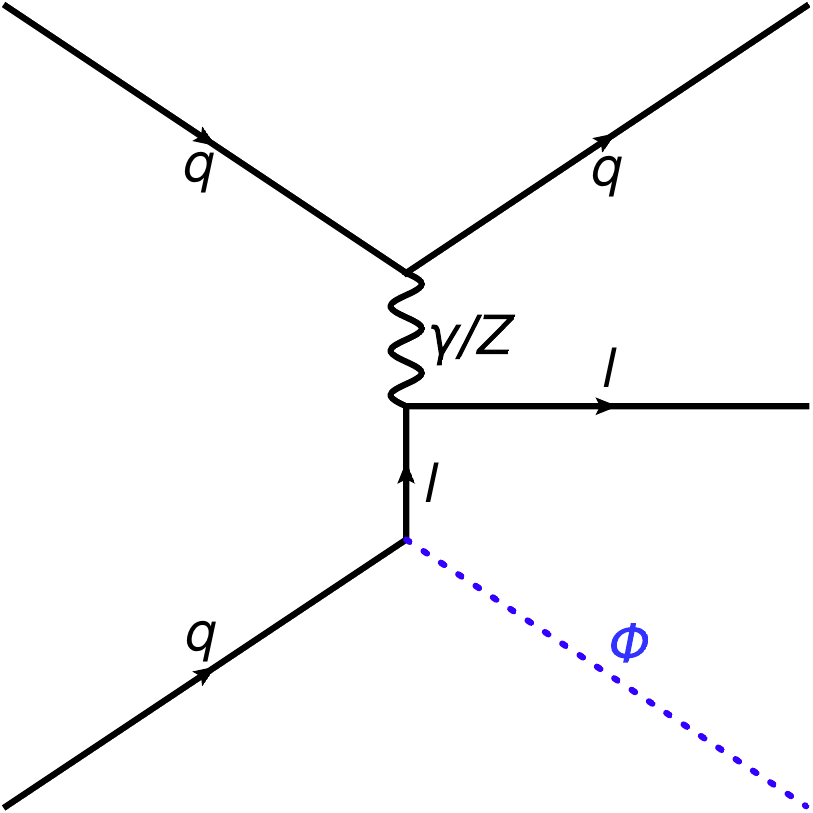}\label{SPqq}}\hfill
\subfloat[(k)]{\includegraphics[width=0.19\textwidth]{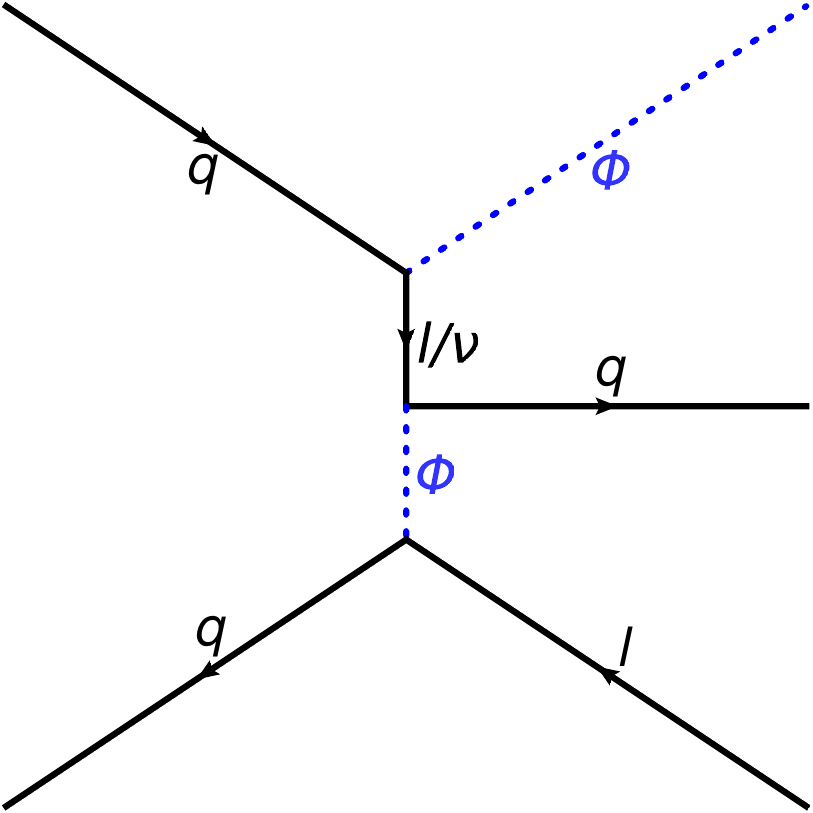}\label{SPSLQ3}}\hfill
\subfloat[(l)]{\includegraphics[width=0.19\textwidth]{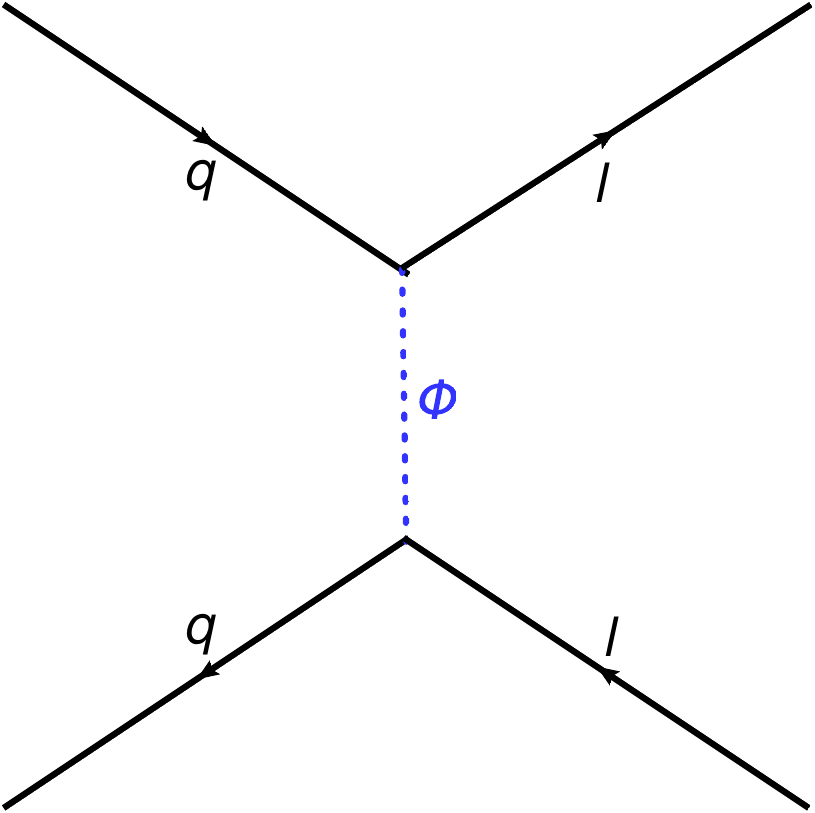}\label{indirect}}
\caption{Representative Feynman diagrams for pair, single and indirect productions of sLQs. A generic sLQ is denoted by $\Phi$.\label{feyn_dgm}}
\end{figure*}
%%%%%%%%%%%%%%%%%%%%%%%%%%%%%%%%%%%%%%%%%%%%%%%%%%%%%%%%%%%%%%%%%%

%%%%%%%%%%%%%%%%%%%%%%%%%%%%%%%%%%%%%%%%%%%%%%%%%%%%%%%%%%%%%%%%%%
\begin{figure*}%[!t]
\centering
\captionsetup[subfigure]{labelformat=empty}
\subfloat[\quad\quad(a)]{\includegraphics[height=6cm,width=6.375cm]{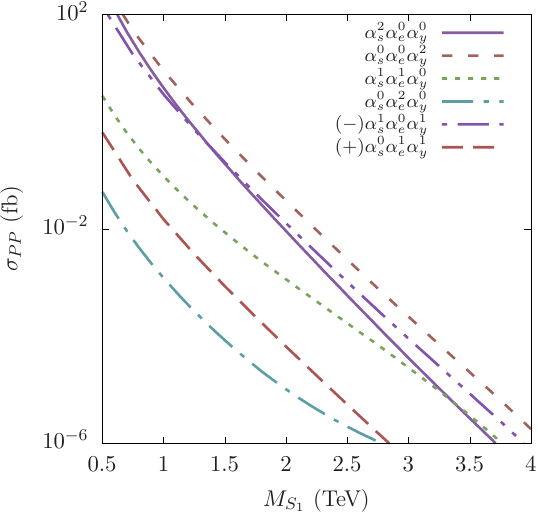}}\hspace{1cm}
\subfloat[\quad\quad(b)]{\includegraphics[height=6cm,width=6.375cm]{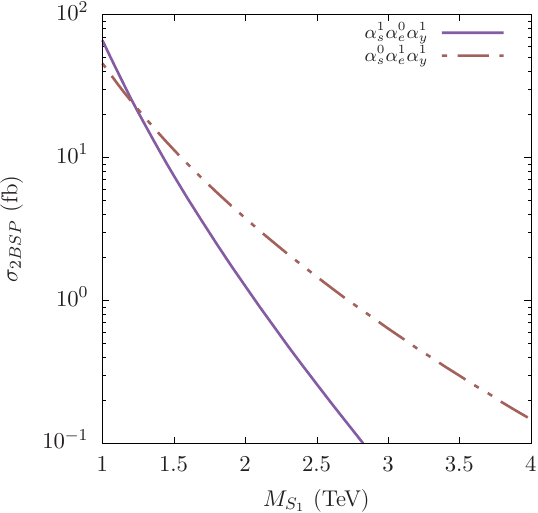}}\\
\subfloat[\quad\quad(c)]{\includegraphics[height=6cm,width=6.375cm]{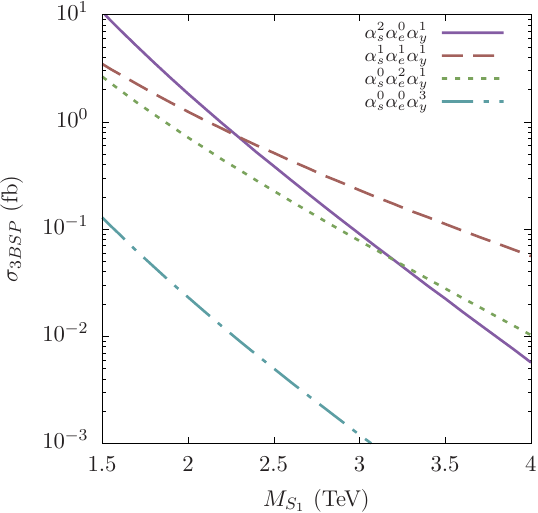}}\hspace{1cm}
\subfloat[\quad\quad(d)]{\includegraphics[height=6cm,width=6.375cm]{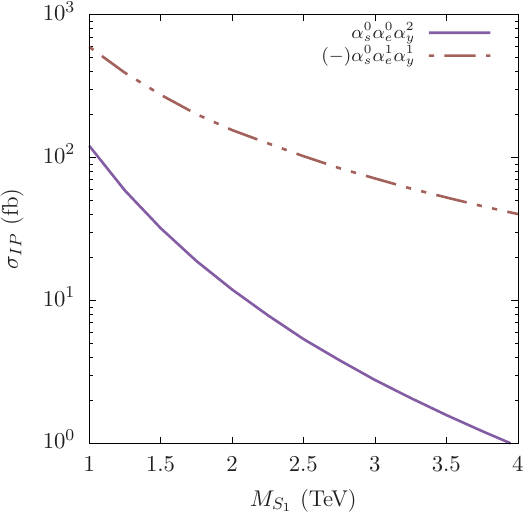}}
\caption{The PP, SP, IP, and II contributions for $S_1$ (subjected to a $M_{\mu^{+}\mu^{-}} \geq 120$ GeV cut) as functions of $M_{S_1}$ for $y^{LL}_{10,12}=1$ at different orders of 
$\alpha_{s}$, $\alpha_{e}$, and $\alpha_{\lambda}$. The constructive/destructive interference contributions are indicated with $(\pm)$ signs in front. 
\label{partonxsec}}
\end{figure*}

\section{Producing scalar leptoquarks at the LHC}
\label{sec:slqprod}
\noindent
In the notation of Ref.~\cite{Dorsner:2016wpm}, all sLQ species except $\overline{S}_1$ (which exclusively couples to right-handed neutrinos), namely, $S_{1}$, $\widetilde{S}_{1}$, $R_{2}$, $\widetilde{R}_{2}$, and $S_{3}$ can directly produce $\ell\ell jj$ final states at the LHC. We display the possible Yukawa interactions of these sLQs with quarks and leptons in Table~\ref{tab:SLQYuk}. Depending on whether the mass eigenstates of the sLQs are aligned with the mass eigenstates of the up-type quarks or the down-type ones, we consider two scenarios---up-aligned and down-aligned.  For simplicity, we do not consider the more general setup where these sLQs are neither up or down-aligned. We also ignore the effect of neutrino mixing and take $U_{\rm PMNS}=1$ as all neutrinos produce missing energy, and the neutrino oscillation length is irrelevant to our analysis.

LQs can be produced at the LHC in two ways---resonantly or nonresonantly. In the resonant modes, the pair and single production (PP and SP) processes produce on-shell LQs, whereas, in the nonresonant production (indirect production---IP), a LQ is exchanged in the $t$-channel. The IP process can interfere (constructively or destructively, depending on the sLQ species) with SM processes. Since all these production topologies can lead to $\ell\ell jj$ final states, we briefly review them. 
\medskip

\noindent
\textbf{Pair production:} Two sLQs (of the same or different species) can be produced resonantly. These decay to produce the $\ell\ell j j$ final states. The leading tree-level contributions to the PP cross section are:

\begin{enumerate}
\item $\mc{O}(\al_s^2\al_e^0\al_y^0)$:\footnote{We use $\al_{y}=y^{2}/4\pi$ where $y$ denotes a sLQ-quark-lepton Yukawa coupling.} The diagrams contributing to this order are purely QCD-mediated; see, e.g., Fig.~\ref{qqlqlq}. Hence, the contribution is model-independent.  
    
\item $\mc{O}(\al_s^0\al_e^2\al_y^0)$: There are $qq$ and $\gm \gm$-initiated processes that depend only on the electric charge of the sLQ [Figs.~\ref{qqzlqlq} and \ref{pplqlq}]. 

\item $\mc{O}(\al_s^0\al_e^0\al_y^2)$: A lepton exchange in the $t$-channel can lead to sLQ pair production---see Fig.~\ref{qqtch}. This purely new-physics (NP) contribution is highly sensitive to the LQ Yukawa couplings and the parton distribution functions (PDFs) of the initial quarks.

\item $\mc{O}(\al_s^1\al_e^0\al_y^1)$: The contribution to the total PP cross section at this order comes from the interference between the diagrams that separately produce $\mc{O}(\al_s^2\al_e^0\al_y^0)$ and $\mc{O}(\al_s^0\al_e^0\al_y^2)$ contributions to the cross section. This interference is destructive in nature for all sLQ species.

\item $\mc{O}(\al_s^0\al_e^1\al_y^1)$: This contribution comes from the interference of the diagrams that separately produce $\mc{O}(\al_s^0\al_e^2\al_y^0)$ and $\mc{O}(\al_s^0\al_e^0\al_y^2)$ contributions to the cross section. Generally, this interference is a minor contribution to the PP cross section. 

\item $\mc{O}(\al_s^1\al_e^1\al_y^0)$: Finally, there is also a QCD-QED mixed order contribution to the PP [see Fig.~\ref{gglqlq3pt}], which, unlike the above two cases, does not come from any interference term (colour conservation stops the $qq \to g \to \ell_q\ell_q$ and $qq \to \gamma/Z \to \ell_q\ell_q$ processes from interfering) but the gluon-photon-sLQ-sLQ term in the kinetic Lagrangian~\cite{Crivellin:2021ejk}. %(see Appendix~\ref{appendix:kinlag}~\cite{Crivellin:2021ejk}), leads to this contribution to the PP cross section [see Fig.~\ref{gglqlq3pt}].

\end{enumerate}
Putting these together, we can make the parameter dependence of the PP cross section explicit as
\begin{align}
\sigma_{\rm PP}(M_{\ell_q},y) =&\ \sigma_{\rm PP}^{200}(M_{\ell_q}) + \sigma_{\rm PP}^{110}(M_{\ell_q}) + \sigma_{\rm PP}^{020}(M_{\ell_q}) \nn \\
&+ y^2\,\overline{\sigma_{\rm PP}^{101}(M_{\ell_q})} +  y^2\, \overline{\sigma_{\rm PP}^{011}(M_{\ell_q})} + y^4\,\overline{\sigma_{\rm PP}^{002}(M_{\ell_q})},
\end{align}
where $\sg_{\rm P}^{ijk}$ denotes the contribution of order $\al_s^i\al_e^j\al_y^k$ to the process ``${\rm P}$'' and the overlines indicate the functions under are evaluated at $y=1$, i.e., $\overline{\sigma^{ijk}_P(M_{\ell_q})}=\sigma^{ijk}_P(M_{\ell_q},y=1)$.
\medskip

\noindent
\textbf{Single production:} The new Yukawa couplings allow a single sLQ to be produced resonantly in association with a lepton and jets ($pp\to \ell_q \ell$ or $pp\to \ell_q \ell j$). From there, decays of the sLQ can give the desired final states. There are two-body single productions (2BSPs) and three-body single productions (3BSPs). As we discuss below, the $\ell\ell jj$ signal will essentially come from the 3BSP processes at the LHC.\\

\noindent
\textbf{2BSP:} There are two leading-order (LO) contributions to this process: 
\begin{enumerate}
\item $\mathcal{O}(\al_s^1\al_e^0\al_y^1)$: There are $qg$-initiated contributions to $pp\to\ell_q\ell$ [shown in  Fig.~\ref{2bspqg}].

\item $\mathcal{O}(\al_s^0\al_e^1\al_y^1)$: Similarly, a $q\gm$-initiated diagram is shown in Fig.~\ref{2bspqgamma}.
\end{enumerate}
\medskip

\noindent
\textbf{3BSP:} There are two types of diagrams that contribute to the 3BSP process. First, when a (hard, i.e., separable) jet radiates off a 2BSP process [i.e., an initial/final state radiation (ISR/FSR) jet], we count it as a 3BSP process:

\begin{enumerate}
\item $\mc{O}(\al_s^{1+(1)}\al_e^0\al_y^1)$:\footnote{The parentheses in the exponents count the number of hard radiations.} A (hard) jet is emitted from the $\mathcal{O}(\al_s^1\al_e^0\al_y^1)$ contribution to the 2BSP process; see, e.g.,  Fig.~\ref{SPqg} where the jet is emitted from the LQ; the jet can also come off the initial partons.

\item$\mc{O}(\al_s^{(1)}\al_e^1\al_y^1)$: An (hard) ISR or FSR from the $\mathcal{O}(\al_s^0\al_e^1\al_y^1)$ contribution to the 2SBP can contribute to this order. See, e.g., Fig.~\ref{SPqp}.

\item[3.]  $\mc{O}(\al_s^0\al_e^{1+(1)}\al_y^1)$: Similarly, when an initial state quark splits into a hard quark (producing a hard ISR) and an electroweak boson, which further interacts with another quark to produce a LQ and a lepton [as in Fig.~\ref{2bspqgamma}], we count it as a separate 3BSP diagram; see Fig.~\ref{SPqq}.
\end{enumerate}
Second, as shown in Ref.~\cite{Mandal:2015vfa} (also see~\cite{Mandal:2012rx}), $pp\to \ell_q \ell j$ contain some genuine NP contributions where the jet is neither an ISR nor a FSR off the $pp\to \ell_q \ell$ processes:  

\begin{enumerate}
\item[4.] $\mc{O}(\al_s^0\al_e^0\al_y^3)$: This contribution comes from another set of $qq$-initiated diagrams that depend only on the LQ Yukawa coupling(s)---a representative Feynman diagram is shown in Fig.~\ref{SPSLQ3}. 

\item[5.]  There is another possibility where the lepton and the jet come from an off-shell sLQ, i.e., $pp\to\ell_q\ell_q^*\to\ell_q\ell j$. Hence, the process contributes at one higher power of $\alpha_y$ than the PP subprocess---$\mc{O}(\al_s^2\al_e^0\al_y^1)$, $\mc{O}(\al_s^0\al_e^2\al_y^1)$,  $\mc{O}(\al_s^0\al_e^0\al_y^3)$, $\mc{O}(\al_s^1\al_e^0\al_y^2)$, $\mc{O}(\al_s^0\al_e^1\al_y^2)$, and $\mc{O}(\al_s^1\al_e^1\al_y^1)$. If the sLQ is far off-shell, it will have different kinematics than that of the PP processes.
\end{enumerate}

%%%%%%%%%%%%%%%%%%%%%%%%%%%%%%%%%%%%%%%%%%%%%%%%%%%%%%%%%%%%%%%%%%
\begin{figure}[t]
\centering
\includegraphics[height=6cm,width=6.375cm]{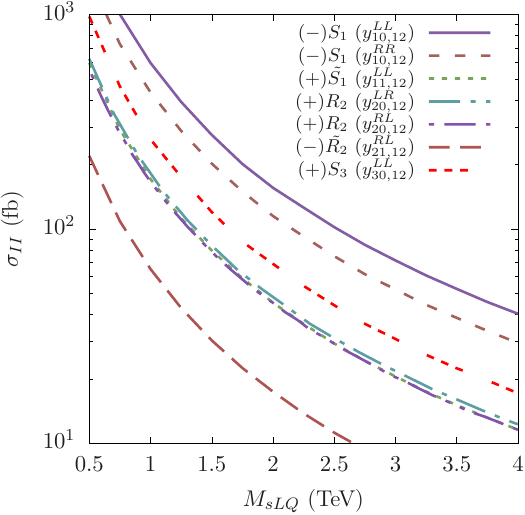}
\caption{Indirect interference contributions for different sLQs.\label{fig:ii}}
\end{figure}
%%%%%%%%%%%%%%%%%%%%%%%%%%%%%%%%%%%%%%%%%%%%%%%%%%%%%%%%%%%%%%%%%%

Estimating the SP contribution to $\ell\ell j j$ states systematically requires some care~\cite{Mandal:2015vfa}. Even though the final state of the 2BSP processes is $\ell\ell j$ ($pp\to \ell_q \ell\to \ell q \ell$) at the Born level, not $\ell\ell jj$, these processes can emit an extra jet while showering and contribute to the $\ell\ell jj$ signals. However, if the radiation is soft/collinear, it will be hard to distinguish it experimentally from the Born-level process. Hence, to avoid confusion, we club such processes where the extra jet is hard with 3BSP. This also agrees with the experimental schemes where jets are identified with reasonably hard $p_T$ cuts. Following Ref.~\cite{Mandal:2015vfa}, we can mathematically express the SP contribution to $\ell\ell j j$ final states with the following equation,
\begin{align}
\label{eq:incluSP}
\sg_{\rm SP}
=&\ \underbrace{\big(\sg^{\rm LO}_{2BB(\ell_q\ell)} + \overbrace{\sg^{\rm virtual}_{(\ell_q\ell)} + \sg_{{2BR}(\ell_q\ell j)}^{\rm soft+collinear}}^{\rm Divergent~terms} + \sg_{{3BNS}(\ell_q\ell j)}^{\rm soft}\big)}_{\rm Negligible~contribution} \nn\\
&\ +\underbrace{\big(\sg_{{2BR}(\ell_q\ell j)}^{\rm hard} + \sg_{{3BNS}(\ell_q\ell j)}^{\rm hard}\big)}_{\rm Main~contribution \ (\approx\ 3BSP)}\ +\ \cdots,
\end{align}
where the suffix $2BB$ indicates two-body Born-level processes (2BSP), $2BR$ denotes 2BSP$+$jet processes and $3BNS$ indicates the NP contribution in 3BSP. In other words, the main contribution to $\ell\ell jj$ signals from the single productions comes from the 3BSP processes. Hence, the parametric dependence of the SP can be made explicit as,
\begin{align}
\sigma_{\rm SP}(M_{\ell_q},y) =&\ y^2\left\{\overline{\sigma_{SP}^{021}(M_{\ell_q})} + \overline{\sigma_{SP}^{201}(M_{\ell_q})} + \overline{\sigma_{SP}^{111}(M_{\ell_q})}\right\} \nn \\
&\ + y^6\overline{\sigma_{SP}^{003}(M_{\ell_q})}.
\end{align}

There is a chance of double counting in generating SP Monte Carlo (MC) events separately as $\ell_q\ell j$ events can also come from PP diagrams as $pp\to\ell_q\ell_q\to\ell_q(\ell j)$. One needs to make sure the lepton and jet in the final state are not coming from an on-shell LQ~\cite{Mandal:2015vfa}.
\medskip

\noindent
\textbf{Indirect production:} The main contribution from this topology to $\ell\ell jj$ final states comes from a single order:

\begin{enumerate}
\item $\mc{O}(\al_s^{(2)}\al_e^0\al_y^2)$: We show an illustrative diagram of the $qq^{(\prime)}\to \ell\ell$ process via a $t$-channel sLQ exchange in Fig.~\ref{indirect}. This process can contribute to the $\ell\ell j j$ signal in the presence of two additional hard radiation jets.
\end{enumerate}

\noindent
\textbf{Indirect interference (II):} The IP process ($qq\to \ell\ell$) interferes with the SM $s$-channel $Z/\gm$-mediated processes with the same initial and final states. We count this contribution separately from the IP as the kinematics of these two are very different.

\begin{enumerate}
\item $\mc{O}(\al_s^{(2)}\al_e^1\al_y^1)$:  Because of the large cross sections of the SM processes, the interference contribution usually plays a major (sometimes, the determining) role in the exclusion limits, especially, in the high-mass and large-coupling(s) region. Its sign depends on the sLQ involved in the diagram; e.g., it is negative for the $S_1$; it is overall positive for the $R_2$; it is negative for the $S_3$ if the initial quarks are up-type but positive if they are down-type; etc.~\cite{Raj:2016aky,Bansal:2018eha}.
\end{enumerate}

\noindent
The total indirect contribution can be expressed as:
\begin{equation}
\sigma_{\rm ind}(M_{\ell_q},y) = y^2\overline{\sigma_{\rm II}^{011}(M_{\ell_q})} + y^4\overline{\sigma_{\rm IP}^{002}(M_{\ell_q})}.
\end{equation}

For an illustration, we plot the cross sections of different orders and topologies for a down-aligned $S_1$ with the coupling $y^{LL}_{10,12}=1$ (see Appendix~\ref{appendix:notation} for the notation) with respect to its mass at the $13$ TeV LHC in Fig.~\ref{partonxsec}. There, we see that as $M_{S_1}$ increases, the photon-initiated contribution to the 3BSP process becomes more important than the gluon-initiated ones due to the lighter $t$-channel exchange (see the corresponding Feynman diagrams in Fig.~\ref{feyn_dgm}). The II contributions for various sLQs are shown in  Fig.~\ref{fig:ii}.

%%%%%%%%%%%%%%%%%%%%%%%%%%%%%%%%%%%%%%%%%%%%%%%%%%%%%%%%%%%%%%%%%%
\begin{table*}
\caption{The cross sections ($\sigma$), the numbers of events ($\mc N$) surviving the $\m\m j j$-selection cuts~\cite{ATLAS:2020dsk} for $\mc L_{exp}= 139$ fb$^{-1}$ of integrated luminosity, and the cut efficiencies [$\varepsilon$, defined in Eq.~\eqref{eq:cuteff}] for three benchmark masses of $S_1$. As in Fig.~\ref{partonxsec}, we have set $y_{10,12}^{LL}=1$ to obtain the numbers.  We see that the importance of the indirect modes increases with the mass of the LQ.\label{tab:table}}
\centering{\small\renewcommand\baselinestretch{1.8}\selectfont
\begin{tabular*}{\textwidth}{c @{\extracolsep{\fill}}rrr rrr rrr rrr}
\hline
\textbf{Mass} ($S_{1}$) & \multicolumn{3}{c}{\textbf{Pair Production}} & \multicolumn{3}{c}{\textbf{Single Production}} & \multicolumn{3}{c}{\textbf{Indirect Production}} & \multicolumn{3}{c}{\textbf{Indirect Interference}} \\ \cline{2-4}\cline{5-7}\cline{8-10}\cline{11-13}
(TeV) & $\sigma_{\rm PP}$ (fb) & $\epsilon_{\rm PP}$ & $\mathcal{N}_{\rm PP}$ & $\sigma_{\rm SP}$ (fb) & $\epsilon_{\rm SP}$ & $\mathcal{N}_{\rm SP}$ & $\sigma_{\rm IP}$ (fb) & $\epsilon_{\rm IP}$ & $\mathcal{N}_{\rm IP}$ & $\sigma_{\rm II}$ (fb) & $\epsilon_{\rm II}$ & $\mathcal{N}_{\rm II}$   \\ \hline\hline
$1.5$ & $1.4 \times 10^{-1}$ & $0.42$ & $8.2$ & $8.4$ & $0.17$ & $198.5$ & $32.0$ & $0.018$ & $111.2$ & $-275.0$ & $0.007$ & $-267.6$ \\ 
$2.5$ &  $6.9 \times 10^{-4}$ & $0.32$ & $3.1\times10^{-2}$& $5.4 \times 10^{-1}$ & $0.13$ & $9.8$ &  $5.4$ & $0.018$ & $13.4$& $-102.0$ & $0.007$ & $-99.2$ \\ 
$4.0$ & $4.6 \times10^{-7}$ &$0.30$ & $2.0 \times10^{-5}$ & $3.4 \times 10^{-2}$ & $0.20$ & $0.7$ & $1.0$ &$0.028$ & $3.7$ & $-40.2$ & $0.010$ & $-55.9$ \\ \hline
\end{tabular*}}
\end{table*}
%%%%%%%%%%%%%%%%%%%%%%%%%%%%%%%%%%%%%%%%%%%%%%%%%%%%%%%%%%%%%%%%%%

\section{Recasting LHC searches}
\label{sec:recast}
\subsection{Direct-search data}
\noindent
We recast the current ATLAS (observed) limits on sLQ pair production in the $\m\m jj$ channel~\cite{ATLAS:2020dsk} by recalculating the signal. We make the signal inclusive by including all the contributions to the $\m\m jj$ final state discussed in the previous section. To recast, we rely on the following relation between the number of observed events at a particular $M_{\ell_q}$ and the signal cross sections:
\begin{align}
    \frac{\mc N_{obs}(M_{\ell_q})}{\mathcal{L}_{exp}}  =&\ \sigma_{obs}(M_{\ell_q})\times\epsilon_{exp}(M_{\ell_q}) \nn\\
    =&\ \sum_{i~\in~{\rm topologies}}\sigma_i(M_{\ell_q},\vec y)\times \beta_i(\vec y)\times\epsilon_{i}(M_{\ell_q}),
\end{align}
where $\ep$'s denote the final selection efficiencies (including the acceptance), $\beta$ the branching ratios (BRs), and $\mc L_{exp}$ the integrated luminosity. In the second line, the sum goes over different topologies: PP, SP, IP, and II. Here, we have assumed the efficiencies to be largely independent of the couplings ($\vec y$), which is a reasonable assumption since the selection cuts are kinematic in nature. 

To estimate the new efficiencies, we implement all the sLQ Lagrangians in \textsc{FeynRules}~\cite{Alloul:2013bka} to get the UFO model files \cite{Degrande:2011ua,Darme:2023jdn} for \textsc{MadGraph5}~\cite{Alwall:2014hca}. We use the \textsc{NNPDF} PDFs~\cite{NNPDF:2021uiq} with dynamical renormalisation and factorisation scales to generate LO events; which are passed through \textsc{Pythia8}~\cite{Bierlich:2022pfr} for showering and hadronisation, and \textsc{Delphes}~\cite{deFavereau:2013fsa} for detector effects. Jets are formed using the anti-$k_{T}$~\cite{Cacciari:2008gp} clustering algorithm in \textsc{FastJet}~\cite{Cacciari:2011ma}. In our analysis, we use LO cross sections for all signal processes except the PP, for which the NLO corrections are known~\cite{Kramer:2004df,Mandal:2015lca,Borschensky:2020hot,Borschensky:2021hbo,Borschensky:2021jyk,Borschensky:2022xsa}. We include a constant $K$ factor of $1.6$ for this process. 

We pass the simulated events through ATLAS selection criteria 
to estimate the efficiencies for different topologies as
\begin{equation}
    \epsilon_{i} = \frac{\mathcal{N}_{i}}{\mathcal{N}_{i}^{gen}}\mbox{~with~} i =\{\mbox{PP, SP, IP, II}\},\label{eq:cuteff}
\end{equation}
where $\mathcal{N}_{i}$ is the number of events surviving the cuts and $\mathcal{N}_{i}^{gen}$ is the total number of generated Monte Carlo events for the $i^{\rm th}$ topology. 
For validation, we reproduced the PP limits in Refs.~\cite{Dorsner:2022ibm} and~\cite{ATLAS:2020dsk} on $\lambda-M_{\rm LQ}$ and $\beta-M_{\rm LQ}$ planes for all LQs.We have validated the model-independent limits on all sLQ masses from Ref.~\cite{Dorsner:2022ibm} within a few GeVs. For an illustration, we list the cross sections, the numbers of events surviving the $\m\m jj$-channel cuts, and the cut efficiencies for three benchmark masses of $S_1$ in Table~\ref{tab:table}. 

Before we move on, we note that in Ref.~\cite{ATLAS:2020dsk}, the observed limits go up to $M_{\ell_q} = 2$ TeV. For our purpose, we have extrapolated the limits beyond $2$ TeV assuming the experimental selection cuts remain the same in this range, i.e., $\mc N_{obs}(M_{\ell_q}> 2~{\rm TeV})=\mc N_{obs}(M_{\ell_q}= 2~{\rm TeV})$.

\subsection{Dilepton-search data}
\noindent 
We also consider the latest dimuon search data from CMS~\cite{CMS:2021ctt} to put bounds on the new Yukawa couplings as functions of sLQ masses. Since the leptons coming from TeV-scale LQ decays or those produced in association with LQs are expected to have high $p_{\rm T}$, they will affect the high-$p_{\rm T}$ tail of the dilepton (in our case, dimuon) distribution in general. Following the method illustrated in Refs.~\cite{Mandal:2018kau,Bhaskar:2021pml} (also see~\cite{Aydemir:2019ynb,Bhaskar:2022vgk}), we perform $\chi^2$ estimations to obtain the $2\sigma$ exclusion limits on $y$'s by fitting the dimuon-$p_T$ distribution for any fixed-$M_{\ell_q}$ hypothesis. We compute the binwise signal efficiencies by passing the events from all the production processes mentioned above through the cuts used in the dimuon-search analysis and then combining them. We note that since there is no restriction on the number of hard jets in this case, a priori, more processes will contribute to the $\m\m$ signal than the $\m\m jj$ one. For example, the 2BSP processes or the $qq\to \m\m$ process via $t$-channel LQ exchange (without the hard radiations), which essentially did not contribute earlier, will contribute to the dilepton signal. We include all contributions systematically.

%%%%%%%%%%%%%%%%%%%%%%%%%%%%%%%%%%%%%%%%%%%%%%%%%%%%%%%%%%%%%%%%%%
\begin{table}[!b]
\caption{Pure QCD [$\mc{O}(\al_s^{2}\al_e^0\al_y^0$] vs pure QCD+QED [$\mc{O}(\al_s^{2}\al_e^0\al_y^0)$ $+$ $\mc{O}(\al_s^0\al_e^2\al_y^0)$ $+$ $\mc{O}(\al_s^{1}\al_e^1\al_y^0)$] mass exclusion limits: Model-independent limits on various sLQ masses (in GeV) with and without the QED contributions. We assume sLQ decays through only one small coupling (shown in parentheses). The limits remain the same for second-generation quarks.\label{tab:SLQYukabcd}}
\centering{\small\renewcommand\baselinestretch{1.5}\selectfont
\begin{tabular*}{\columnwidth}{l @{\extracolsep{\fill}}cc}
\hline
Model & QCD & QCD+QED  \\ \hline\hline
$S_1 (y^{LL}_{10,12})$ &$1418$ & $1423$  \\ 
$S_1 (y^{RR}_{10,12})$  & $1733$ & $1741$  \\ 
$\tilde{S}_1 (y^{RR}_{11,12})$ & $1733$ & $1854$  \\ 
$R_2 (y^{LR}_{20,12})$ & $1852$ &  $2012$ \\ 
$R_2 (y^{RL}_{20,12})$& $1733$ &   $1917$\\ 
$\tilde{R}_2 (y^{RL}_{21,12})$ & $1733$ & $1767$  \\ 
$S_3 (y^{LL}_{30,12})$ & $1772$  & $1882$  \\ \hline
\end{tabular*}}
\end{table}
%%%%%%%%%%%%%%%%%%%%%%%%%%%%%%%%%%%%%%%%%%%%%%%%%%%%%%%%%%%%%%%%%%

%%%%%%%%%%%%%%%%%%%%%%%%%%%%%%%%%%%%%%%%%%
\begin{figure*}
\centering
\captionsetup[subfigure]{labelformat=empty}
\subfloat[\quad\quad(a)]{\includegraphics[height=4.5cm,width=4.5cm]{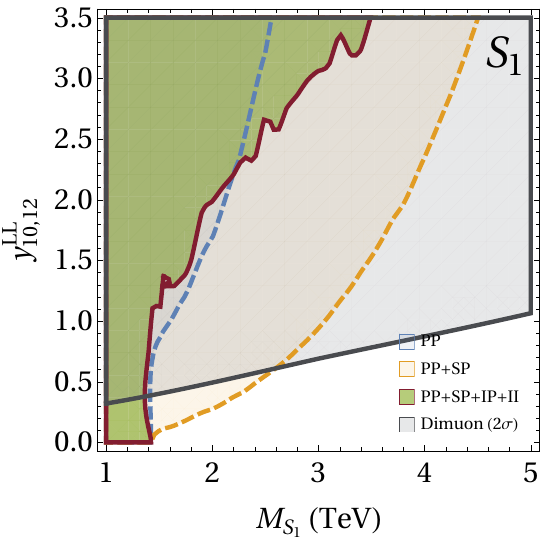}\label{fig:el1s1ll12}}\hfill
\subfloat[\quad\quad(b)]{\includegraphics[height=4.5cm,width=4.5cm]{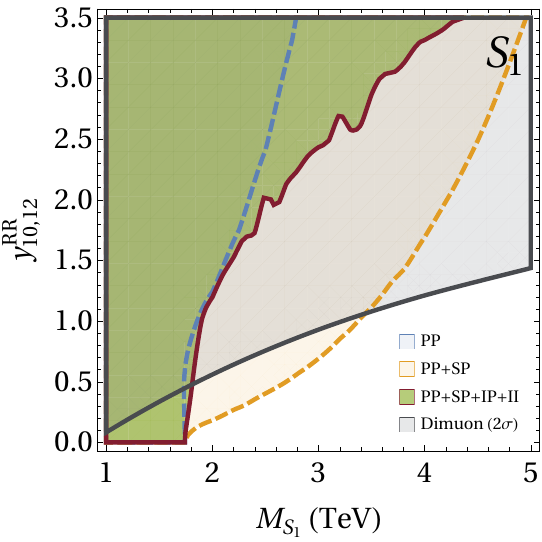}\label{fig:el1s1rr12}}\hfill
\subfloat[\quad\quad(c)]{\includegraphics[height=4.5cm,width=4.5cm]{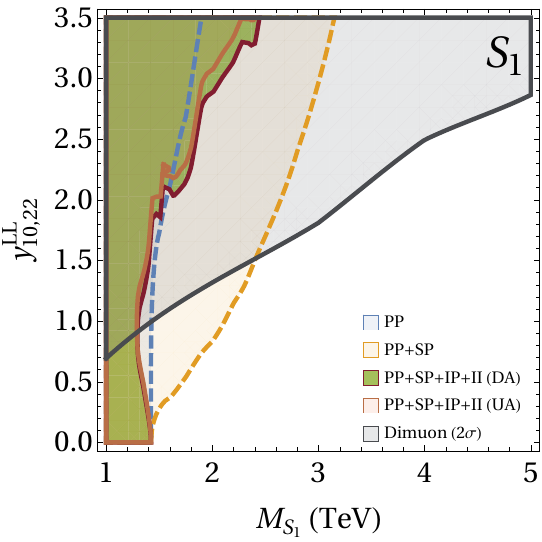}\label{fig:el1s1ll22}}\hfill
\subfloat[\quad\quad(d)]{\includegraphics[height=4.5cm,width=4.5cm]{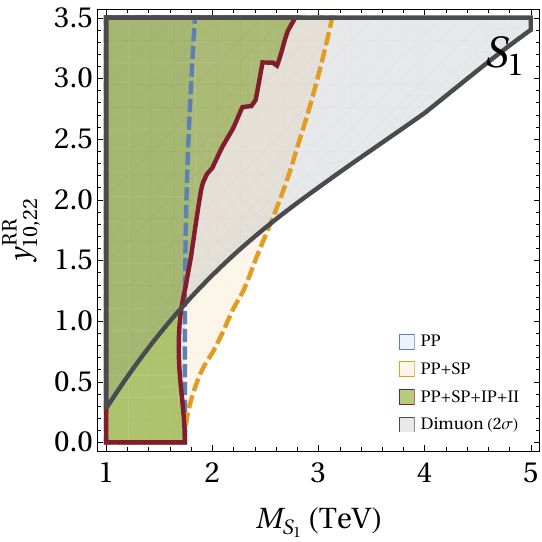}\label{fig:el1s1rr22}}\\
\subfloat[\quad\quad(e)]{\includegraphics[height=4.5cm,width=4.5cm]{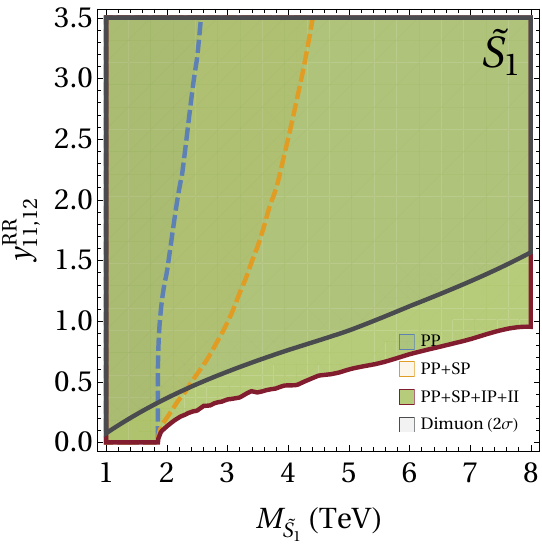}\label{fig:el1s1trr12}}\hfill
\subfloat[\quad\quad(f)]{\includegraphics[height=4.5cm,width=4.5cm]{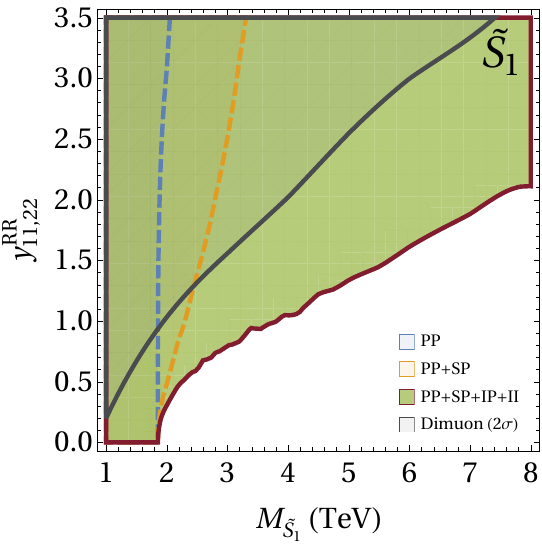}\label{fig:el1s1trr22}}\hfill
\subfloat[\quad\quad(g)]{\includegraphics[height=4.5cm,width=4.5cm]{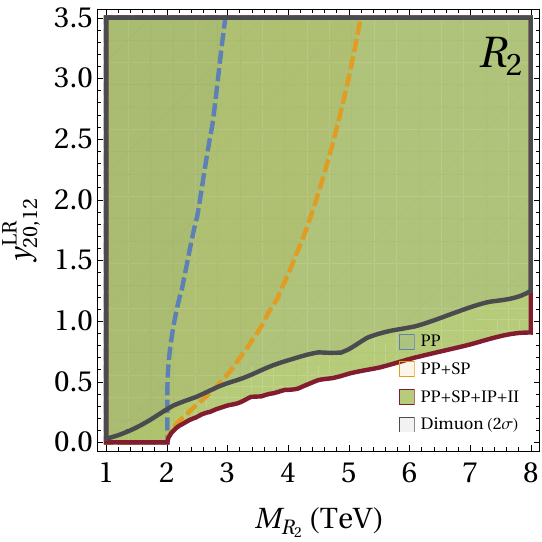}\label{fig:el1r2lr12}}\hfill
\subfloat[\quad\quad(h)]{\includegraphics[height=4.5cm,width=4.5cm]{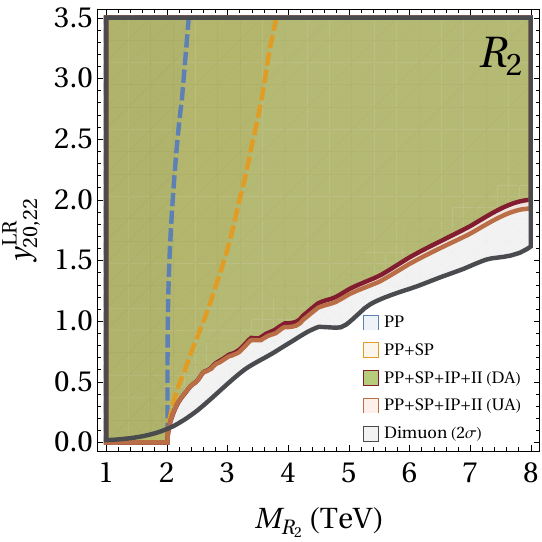}\label{fig:el1r2lr22}}\\
\subfloat[\quad\quad(i)]{\includegraphics[height=4.5cm,width=4.5cm]{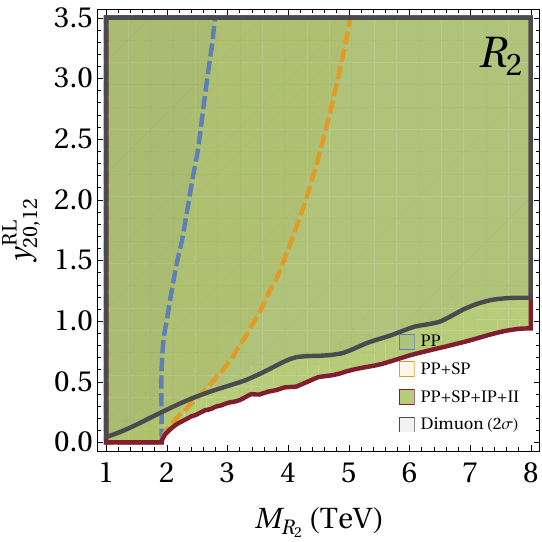}\label{fig:el1r2rl12}}\hfill
\subfloat[\quad\quad(j)]{\includegraphics[height=4.5cm,width=4.5cm]{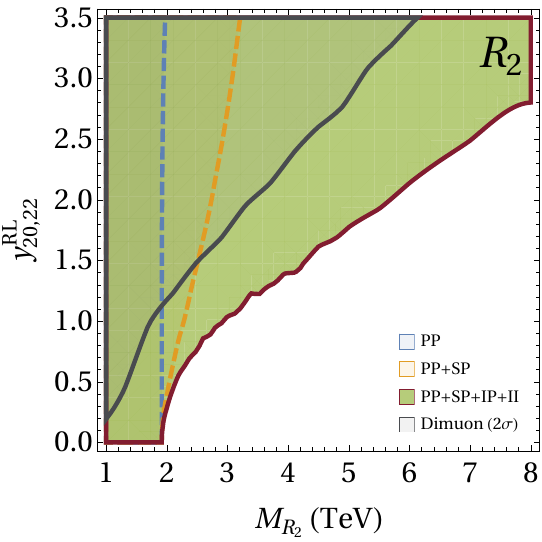}\label{fig:el1r2rl22}}\hfill
\subfloat[\quad\quad(k)]{\includegraphics[height=4.5cm,width=4.5cm]{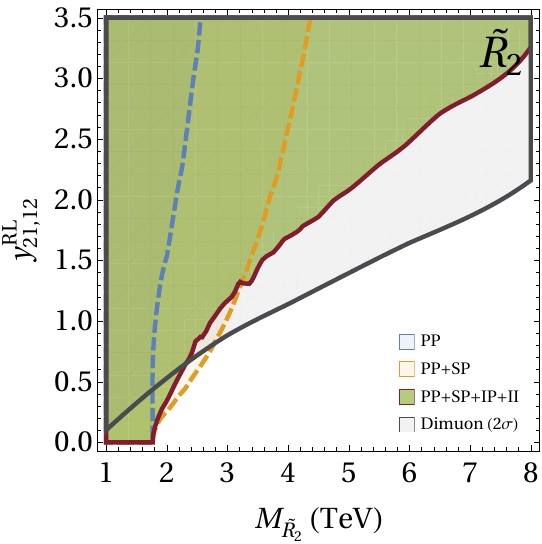}\label{fig:el1r2tr12}}\hfill
\subfloat[\quad\quad(l)]{\includegraphics[height=4.5cm,width=4.5cm]{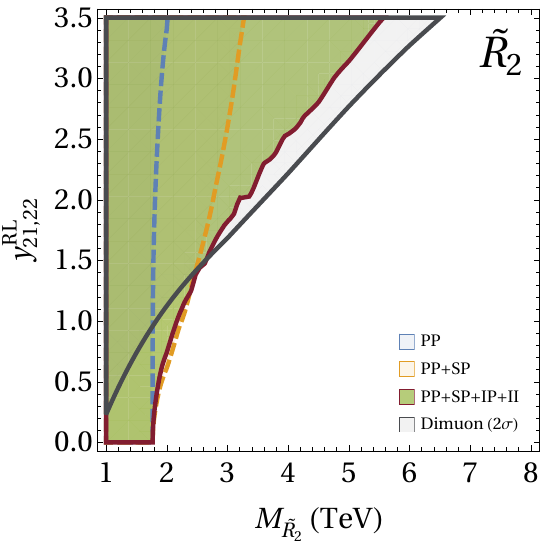}\label{fig:el1r2tr22}}\\
\subfloat[\quad\quad(m)]{\includegraphics[height=4.5cm,width=4.5cm]{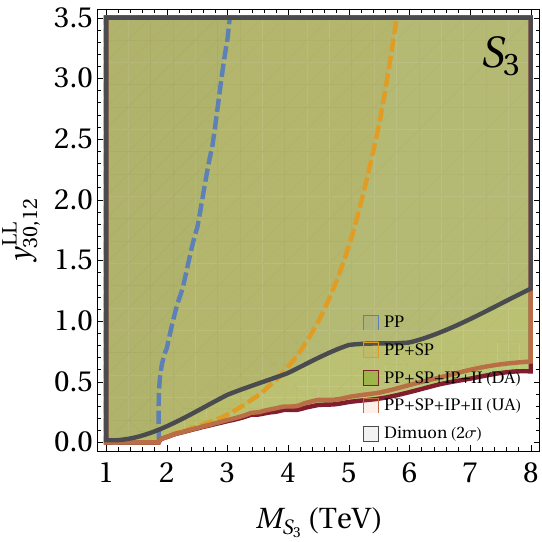}\label{fig:el1s3l12}}\hspace{0.1cm}
\subfloat[\quad\quad(n)]{\includegraphics[height=4.5cm,width=4.5cm]{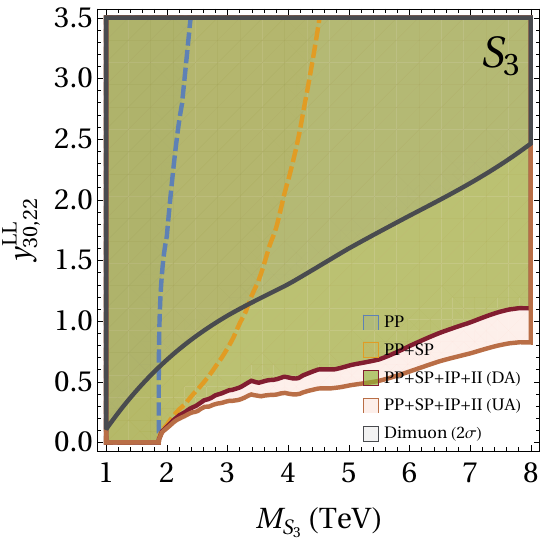}\label{fig:el1s3l22}}\hfill~
\caption{Exclusion limits on all sLQs when only one Yukawa coupling is nonzero (see Appendix~\ref{appendix:notation} for the notation). The contributions from the  PP, PP+SP, and PP+SP+IP+II processes are separately marked; UA and DA indicate up and down-aligned scenarios, respectively (see Table~\ref{tab:SLQYuk}).\label{fig:ELOneCoup}}
\end{figure*}
%%%%%%%%%%%%%%%%%%%%%%%%%%%%%%%%%%%%%%%%%%%%%%%%%

%%%%%%%%%%%%%%%%%%%%%%%%%%%%%%%%%%%%%%%%%%
\begin{figure*}
\centering
\captionsetup[subfigure]{labelformat=empty}
\subfloat[\quad\quad(a)]{\includegraphics[height=4.5cm,width=4.5cm]{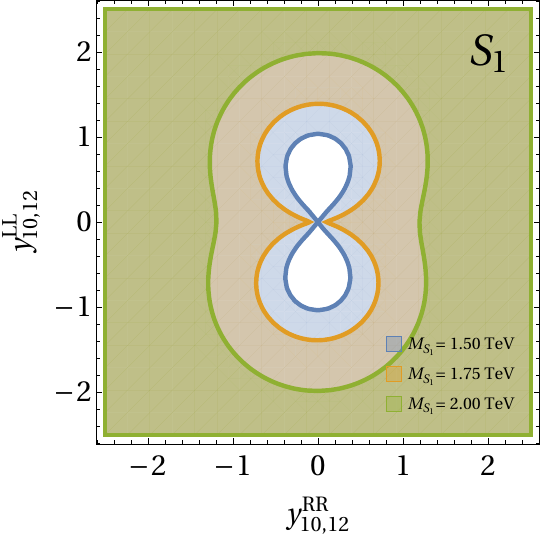}\label{fig:el2s1l12r12}}\hfill
\subfloat[\quad\quad(b)]{\includegraphics[height=4.5cm,width=4.5cm]{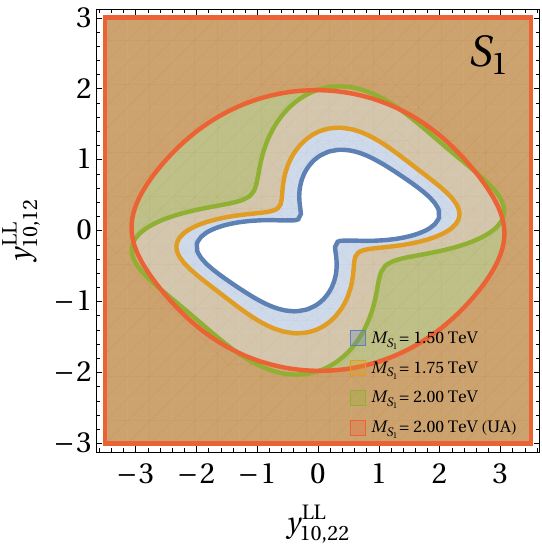}\label{fig:el2s1l12l22}}\hfill
\subfloat[\quad\quad(c)]{\includegraphics[height=4.5cm,width=4.5cm]{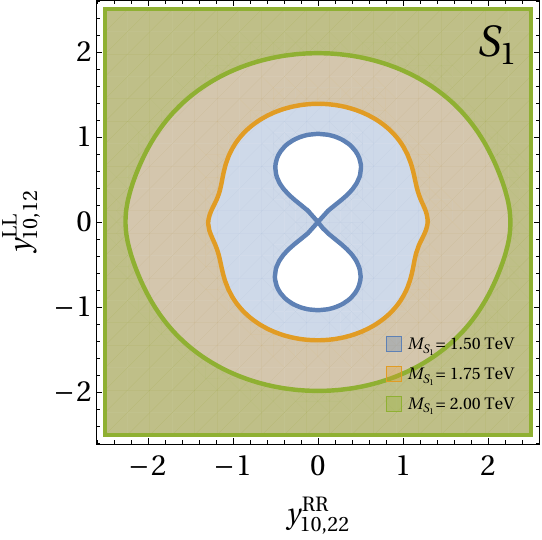}\label{fig:el2s1l12r22}}\hfill
\subfloat[\quad\quad(d)]{\includegraphics[height=4.5cm,width=4.5cm]{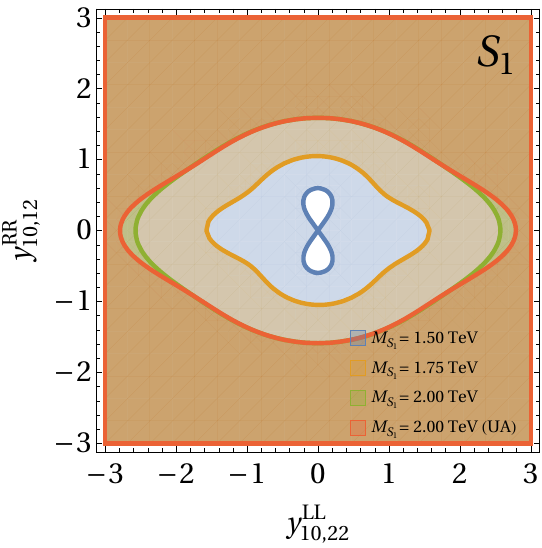}\label{fig:el2s1r12l22}}\\
\subfloat[\quad\quad(e)]{\includegraphics[height=4.5cm,width=4.5cm]{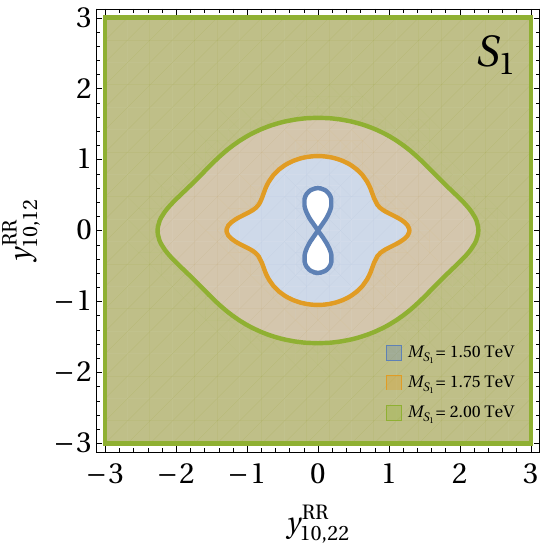}\label{fig:el2s1r12r22}}\hfill
\subfloat[\quad\quad(f)]{\includegraphics[height=4.5cm,width=4.5cm]{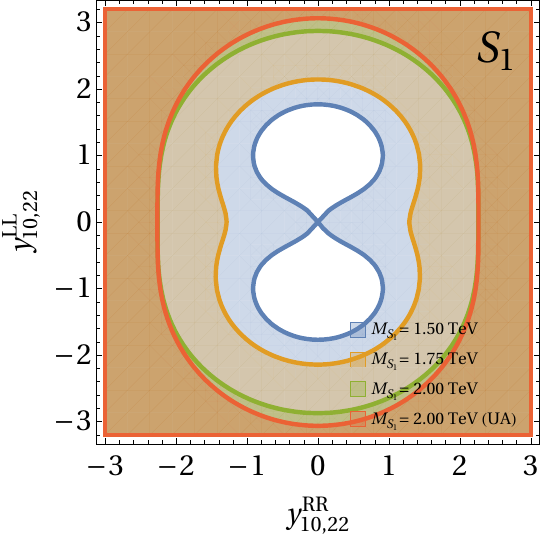}\label{fig:el2s1r12l22}}\hfill
\subfloat[\quad\quad(g)]{\includegraphics[height=4.5cm,width=4.5cm]{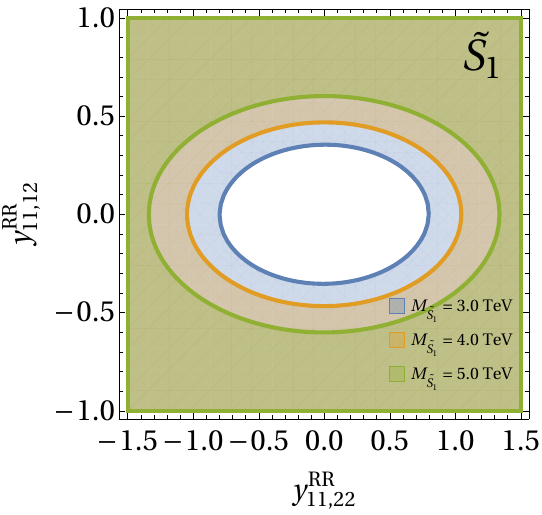}\label{fig:el2s1tr12r22}}\hfill
\subfloat[\quad\quad(h)]{\includegraphics[height=4.5cm,width=4.5cm]{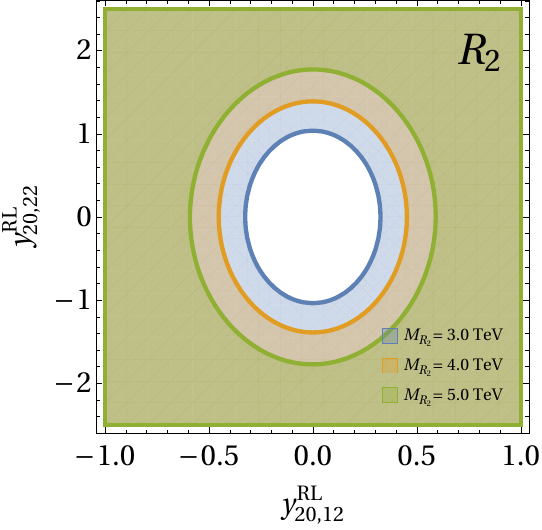}\label{fig:el2r2r22r12}}\\
\subfloat[\quad\quad(i)]{\includegraphics[height=4.5cm,width=4.5cm]{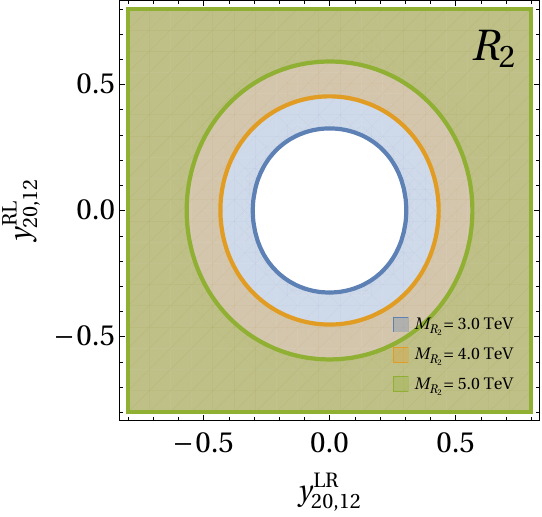}\label{fig:el2r2r12l12}}\hfill
\subfloat[\quad\quad(j)]{\includegraphics[height=4.5cm,width=4.5cm]{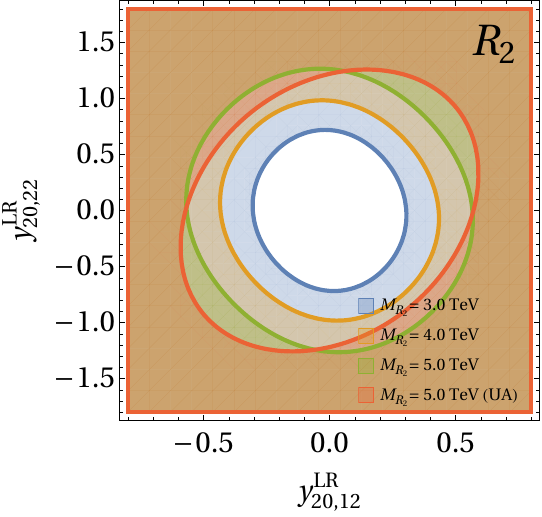}\label{fig:el2r2l22l12}}\hfill
\subfloat[\quad\quad(k)]{\includegraphics[height=4.5cm,width=4.5cm]{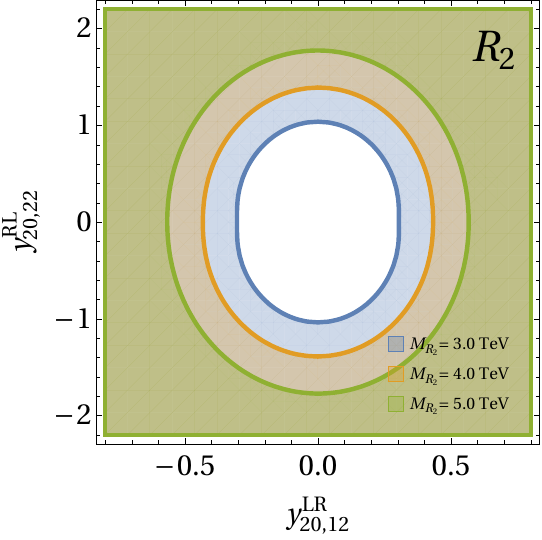}\label{fig:el2r2r22l12}}\hfill
\subfloat[\quad\quad(l)]{\includegraphics[height=4.5cm,width=4.5cm]{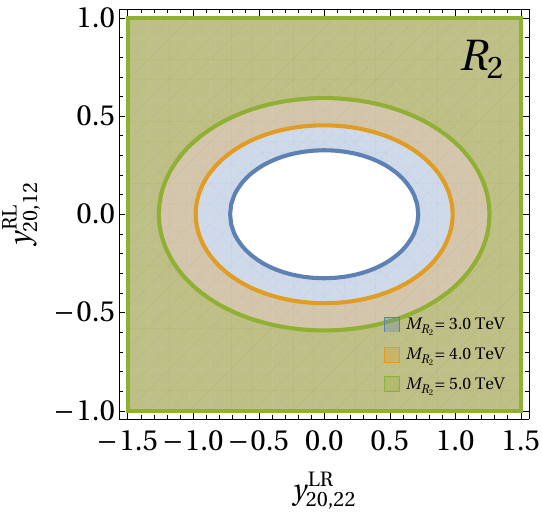}\label{fig:el2r2r12l22}}\\
\subfloat[\quad\quad(m)]{\includegraphics[height=4.5cm,width=4.5cm]{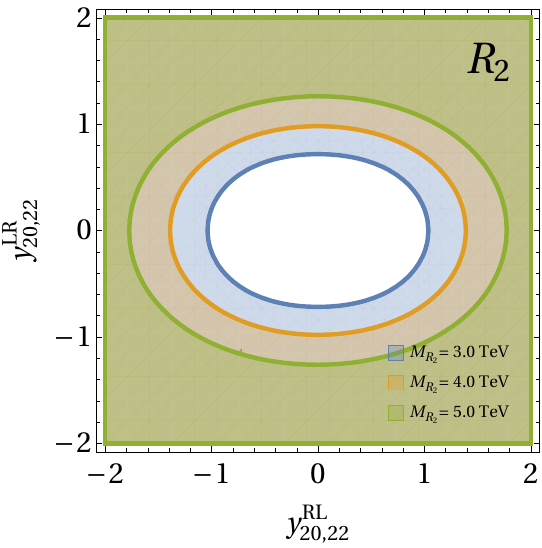}\label{fig:el2r2r22l22}}\hspace{0.1cm}
\subfloat[\quad\quad(n)]{\includegraphics[height=4.5cm,width=4.5cm]{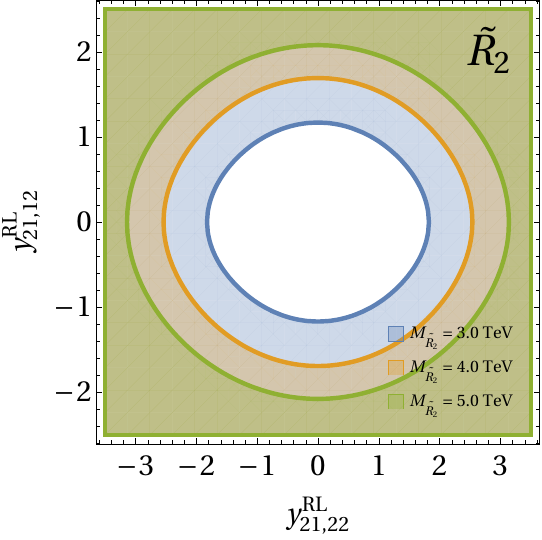}\label{fig:el2r2tr12r22}}\hspace{0.1cm}
\subfloat[\quad\quad(o)]{\includegraphics[height=4.5cm,width=4.5cm]{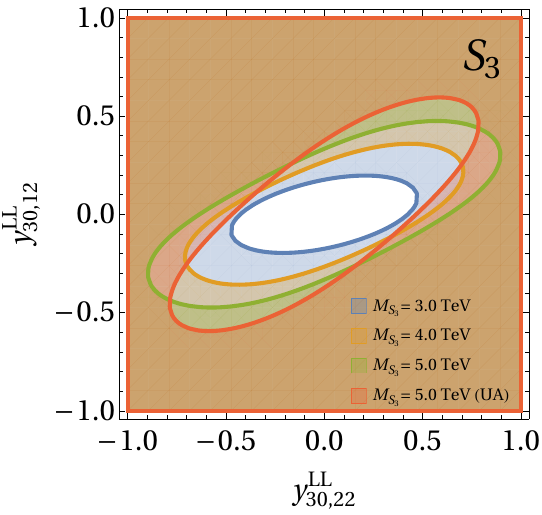}\label{fig:el2s3l12l22}}~\hfill~
\caption{Same as Fig.~\ref{fig:ELOneCoup}, except here, two Yukawa couplings are nonzero.  \label{fig:ELTwoCoup}}
\end{figure*}
%%%%%%%%%%%%%%%%%%%%%%%%%%%%%%%%%%%%%%%%%%%%%%%%%
%%%%%%%%%%%%%%%%%%%%%%%%%%%%%%%%%%%%%%%%%%
\begin{figure*}
\centering
\captionsetup[subfigure]{labelformat=empty}
\subfloat[\quad\quad(a)]{\includegraphics[height=4.5cm,width=4.5cm]{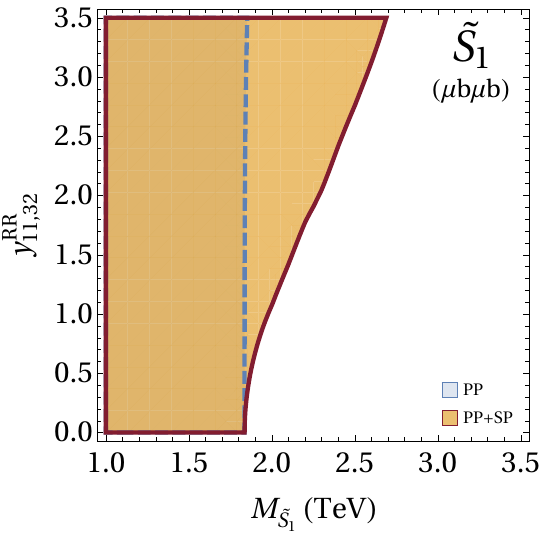}\label{fig:el1s1tr32}}\hfill
\subfloat[\quad\quad(b)]{\includegraphics[height=4.5cm,width=4.5cm]{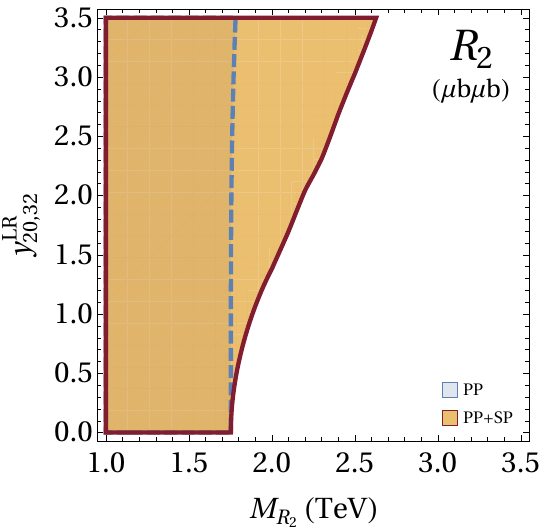}\label{fig:el1r2lr32}}\hfill
\subfloat[\quad\quad(c)]{\includegraphics[height=4.5cm,width=4.5cm]{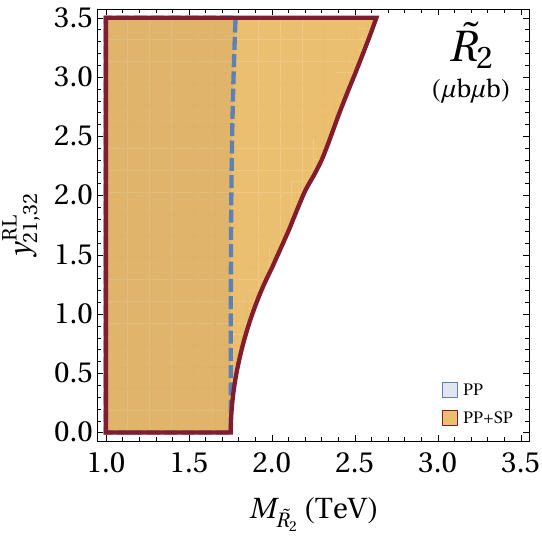}\label{fig:el1r2tr32}}\hfill
\subfloat[\quad\quad(d)]{\includegraphics[height=4.5cm,width=4.5cm]{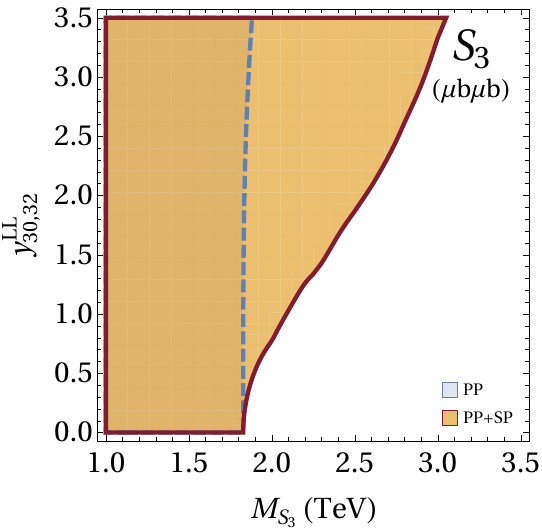}\label{fig:el1s3l32}}\\
\subfloat[\quad\quad(e)]{\includegraphics[height=4.5cm,width=4.5cm]{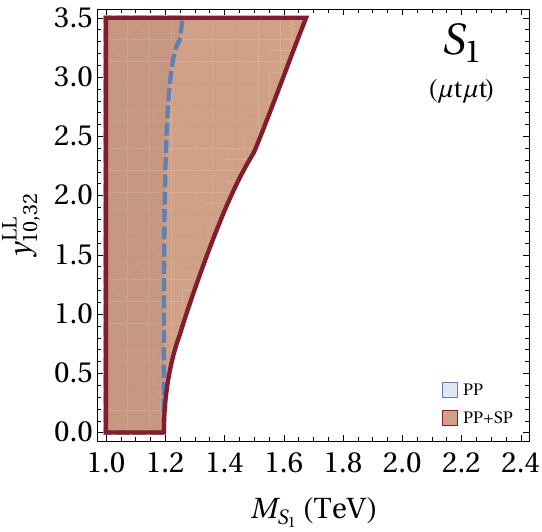}\label{fig:el1s1l32}}\hfill
\subfloat[\quad\quad(f)]{\includegraphics[height=4.5cm,width=4.5cm]{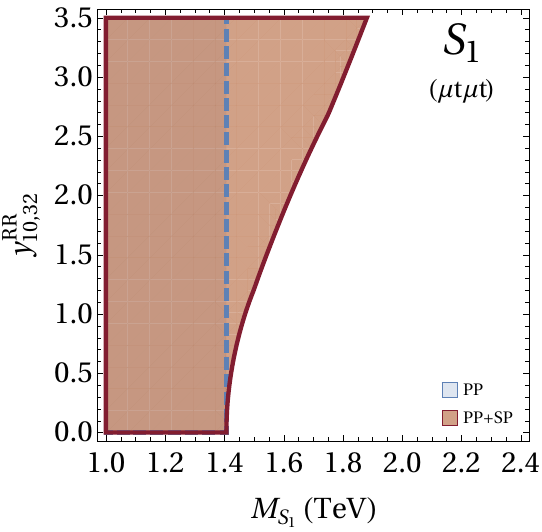}\label{fig:el1s1r32}}\hfill
\subfloat[\quad\quad(g)]{\includegraphics[height=4.5cm,width=4.5cm]{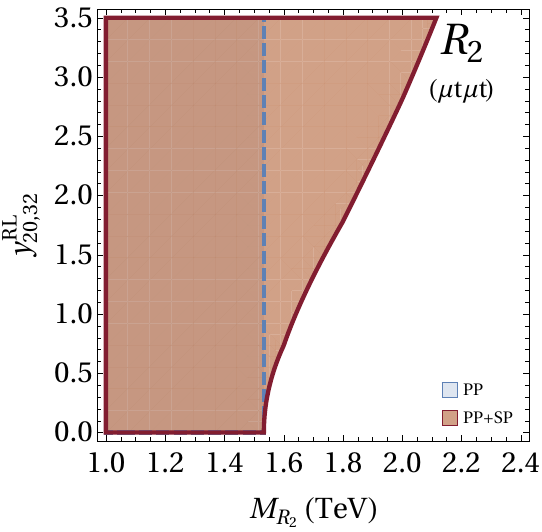}\label{fig:EL_YRL32_DA}}
\subfloat[\quad\quad(h)]{\includegraphics[height=4.5cm,width=4.5cm]{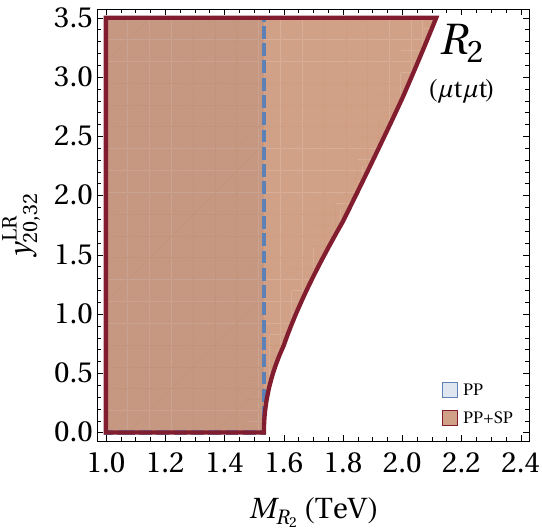}\label{fig:YLR23_ExLim}}\\
\subfloat[\quad\quad(i)]{\includegraphics[height=4.5cm,width=4.5cm]{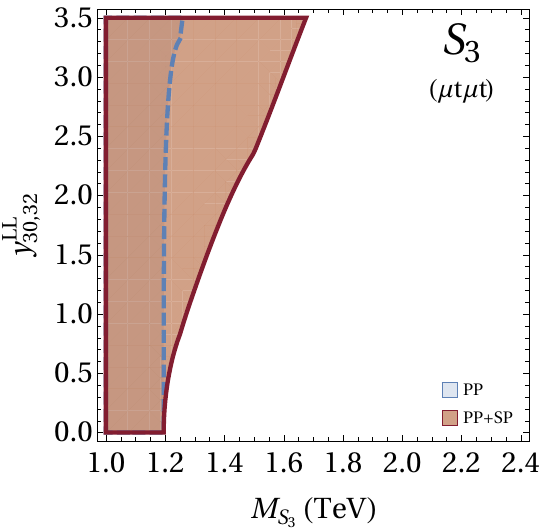}\label{fig:el1s3lmut32}}\hfill
\hfill~
\caption{Single-coupling exclusion limits when the quark involved is a third-generation quark.  
\label{fig:ELOneCoup3rdgen} }
\end{figure*}
%%%%%%%%%%%%%%%%%%%%%%%%%%%%%%%%%%%%%%%%%%%%%%%%%

\section{Exclusion limits}\label{sec:el}
\noindent
We summarise the coupling-independent exclusion limits on the sLQs obtained from the $\mu \mu j j $ data in Table~\ref{tab:SLQYukabcd}. The differences between the limits come from the differences in the BRs.

\subsection{The singlets ($S_1$ and $\widetilde{S_1}$)}
\noindent
The $S_1$ LQ has four Yukawa couplings through which it can decay to $\m j$ final states: $y^{LL}_{10,12}$, $y^{RR}_{10,12}$, $y^{LL}_{10,22}$, and $y^{RR}_{10,22}$. We show the recast ($95\%$ CL) exclusion limits on these couplings separately in Figs.~\ref{fig:el1s1ll12}--\ref{fig:el1s1rr22}. In these plots, we show how the limits change if we consider only the PP or PP+SP or PP+SP+IP+II processes. We also show the $2\sigma$ limits from the high-$p_T$ dimuon tail. We see that for $S_1$, the dimuon limits are the hardest. There are some essential points to note here. First, the indirect interference is destructive in the case of the $S_1$, making the PP+SP+IP+II limits weaker than the PP+SP ones but, at the same time, making the dimuon limits stronger. Second, bigger regions are excluded when $S_1$ mainly couples to the first-generation quarks as compared to the second-generation quarks because of the difference between the PDFs of valence and sea quarks. Third, for the left-handed couplings, BR$(S_1\to \mu j)=50$\% but it is 100\% for the right-handed couplings. As a result, for the same quark generation, the resonant (PP and PP+SP) limits are stronger for the right-handed couplings than for the left-handed ones. Finally, as indicated earlier, the direct limits on the left-handed couplings vary depending on whether the $S_1$ is up-aligned or down-aligned. However, the difference is visible only in the case of the $y^{LL}_{10,22}$ coupling [see Fig.~\ref{fig:el1s1ll22}], i.e., when $S_1$ couples mainly to the second-generation quarks. Because of the smaller second-generation quark PDFs, the relative difference is only marginal for $y^{LL}_{10,12}$. 

Limits on these couplings taken two at a time are shown in Figs.~\ref{fig:el2s1l12r12}--\ref{fig:el2s1r12r22} for different $M_{S_1}$ values. Like the one-coupling plots, regions excluded by the $\m\m jj$ data are shown with colours.

For $\widetilde{S}_1$, we plot the exclusion limits on $y^{RR}_{11,12}$ and $y^{RR}_{11,22}$ on the mass-coupling planes in Figs.~\ref{fig:el1s1trr12} and \ref{fig:el1s1trr22}, respectively. In this case, the interference is positive, making the combined direct limits (PP+SP+IP+II) the strongest and the dimuon limits relatively weaker. The simultaneous limits on these two couplings are shown in Fig.~\ref{fig:el2s1tr12r22}.

%%%%%%%%%%%%%%%%%%%%%%%%%%%%%%%%%%%%%%%%%%%%%%%%%
\begin{figure*}
\centering
\includegraphics[height=6cm,width=6.375cm]{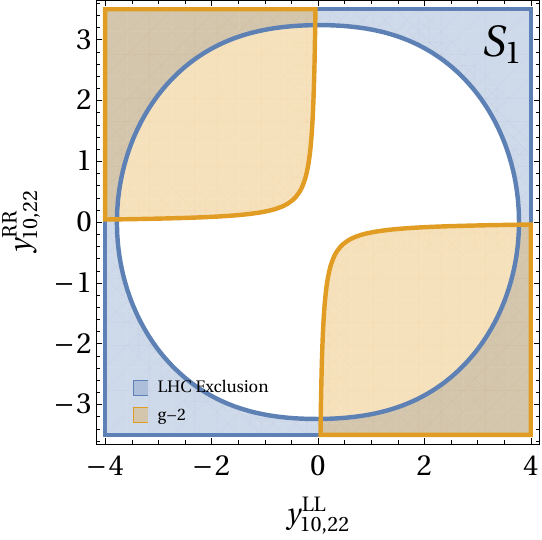}\hspace{1cm}
\includegraphics[height=6cm,width=6.375cm]{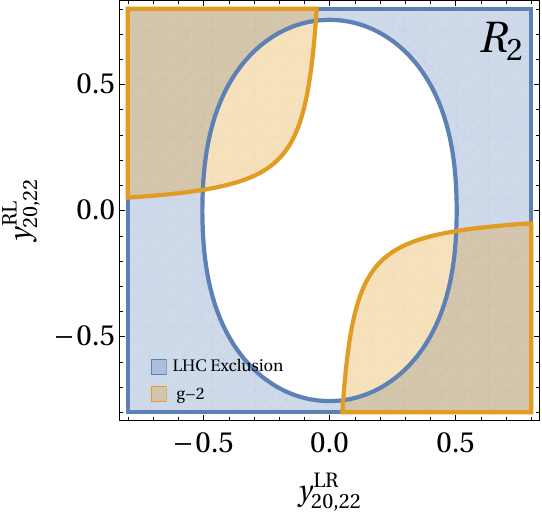}
\caption{The LHC data excludes parts of the parameter space (blue) 
 where $2.5$-TeV sLQs can explain the muon $g-2$ anomaly. \label{fig:gm2}}
\end{figure*}
%%%%%%%%%%%%%%%%%%%%%%%%%%%%%%%%%%%%%%%%%%%%%%%%%
%%%%%%%%%%%%%%%%%%%%%%%%%%%%%%%%%%%%%%%%%%%%%%%%%%%%%%%%%%%%%%%%%%
\begin{table}[!t]
\caption{Extrapolated maximum mass exclusion limit (TeV) for perturbative $y \approx \sqrt{4\pi}$ for different sLQs. To obtain these, we have assumed only one coupling is nonzero at a time.\label{tab:PertLim}}
\centering{\small\renewcommand\baselinestretch{1.6}\selectfont
\begin{tabular*}{\columnwidth}{l @{\extracolsep{\fill}}crcrcrr}
\hline
\multirow{2}{*}{sLQ}& \multirow{2}{*}{$y$} & \multirow{2}{*}{Limit} &\multirow{2}{*}{$y$} & \multirow{2}{*}{Limit} &\multirow{2}{*}{$y$} & Limit&  Limit\\
&&&&&& ($\mu b\m b$) &  ($\mu t \m t$) \\\hline\hline
\multirow{2}{*}{ $S_1$} &  $y^{LL}_{10,12}$& $3.5$  & $y^{LL}_{10,22}$ & $2.5$ & $y^{LL}_{10,32}$ & --- & $1.6$\\ 
 & $y^{RR}_{10,12}$ & $4.4$ & $y^{RR}_{10,22}$ & $2.8$ & $y^{RR}_{10,32}$ & --- & $1.9$ \\\hline 
$\tilde{S_1}$        & $y^{RR}_{11,12}$ & $27.0$  & $y^{RR}_{11,22}$  &  $12.0$ & $y^{RR}_{11,32}$ & $2.7$ & --- \\\hline
\multirow{2}{*}{$R_2$}     &  $y^{LR}_{20,12}$& $28.0$ & $y^{LR}_{20,22}$ & $12.0$  & $y^{LR}_{20,32}$ & $2.6$ & $2.3$\\ 
& $y^{RL}_{20,12}$& $27.0$ & $y^{RL}_{20,22}$ &$9.0$ & $y^{RL}_{20,32}$ & --- & $2.1$ \\\hline 
$\tilde{R_2}$      & $y^{RL}_{21,12}$ & $8.2$ & $y^{RL}_{21,22}$ & $5.6$  & $y^{RL}_{21,32}$ & $2.6$ & --- \\\hline  
$S_3$    & $y^{LL}_{30,12}$ & $33.0$  & $y^{LL}_{30,22}$ &  $18.0$ & $y^{LL}_{30,32}$ & $3.1$ & $1.6$\\ \hline
\end{tabular*}}
\end{table}
%%%%%%%%%%%%%%%%%%%%%%%%%%%%%%%%%%%%%%%%%%%%%%%%%%%%%%%%%%%%%%%%%%

\subsection{The doublets ($R_2$ and $\widetilde{R}_2$)}
\noindent
For $R_2$, we show the separate limits on $y^{LR}_{20,12}$, $y^{LR}_{20,22}$, $y^{RL}_{20,12}$, and $y^{RL}_{20,22}$ on mass-coupling planes in Figs.~\ref{fig:el1r2lr12}--\ref{fig:el1r2rl22}, respectively. The interference is positive in the case of $R_2$ and, with the $y^{LR}_{20,ij}$ couplings, both its components ($R_2^{5/3}$ and $R_2^{2/3}$) can significantly contribute to the $\m\m (jj)$ data, making the recast limits stronger than the singlet cases in general. The two-coupling bounds on $R_2$ are shown in Figs.~\ref{fig:el2r2r22r12}--\ref{fig:el2r2r22l22}.

For $\widetilde{R}_2$, the limits on $y^{RL}_{21,12}$ and $y^{RL}_{21,22}$ are shown in Figs.~\ref{fig:el1r2tr12} and~\ref{fig:el1r2tr22}, respectively. The simultaneous limits on these two couplings for different $M_{\widetilde{R_2}}$ are shown in Fig.~\ref{fig:el2r2tr12r22}.

One can also make similar observations for the doublets as we did in the case of $S_1$ (e.g., the difference coming from first and second-generation PDFs, etc.). However, there is a difference---if only one (muon) coupling is nonzero, BR$(R_2^{n}/\widetilde{R}_2^{5/3}\to \m j)=100\%$ (where $n=5/3$ or $2/3$), irrespective of the chiralities of the quark and the lepton.

\subsection{The triplet $S_3$}
\noindent
There are only two couplings for $S_3$ to contribute to $\m\m(jj)$ final states: $y^{LL}_{30,12}$ and $y^{LL}_{30,22}$. We show the limits on the couplings separately in Figs.~\ref{fig:el1s3l12} and~\ref{fig:el1s3l22}, and simultaneously in Fig.~\ref{fig:el2s3l12l22}.

\subsection{Heavy jets: The third-generation quarks}
\noindent
So far, we have implicitly assumed that the jets in the direct searches are light. However, the ATLAS $\m\m jj$ search allows the jets to be $b$ jets and puts bound on the $\m b \m b$ final state. We recast the $\m b \m b$ mass limits to draw bounds on the sLQ couplings with the bottom quark and the muon ($y_{32}$). For completeness, we also recast the $\mu t \mu t$ bounds obtained by CMS~\cite{CMS:2022nty}. We show the limits on these couplings for various sLQs in Fig.~\ref{fig:ELOneCoup3rdgen}. In these cases, we do not show the relatively minor indirect or interference contributions to the limits. These contributions are small because the third-generation quarks have small/vanishing PDFs and they are less likely to appear as radiation.

\subsection{Illustration: The muon $g-2$ anomaly and the LHC limits}
\noindent
As mentioned in the introduction, one of the major reasons for the recent popularity of sLQs in the literature is their ability to explain experimental anomalies like the one seen in muon $g-2$. In Fig.~\ref{fig:gm2}, we draw the LHC bounds on two sLQs ($S_1$ and $R_2$) known to address the muon $g-2$ anomaly to illustrate how the current LHC $\m\m jj$ data impinge on the relevant parameter space. We have used the expressions available in Refs.~\cite{Dorsner:2016wpm,De:2023acg} to obtain the $g-2$ parameter regions in these plots. One can draw similar comparisons with sLQ-parameter spaces relevant to other anomalies as well.

\subsection{Extrapolating the limits}
\noindent
From the one-coupling plots in Figs.~\ref{fig:ELOneCoup} and~\ref{fig:ELOneCoup3rdgen}, we see that the mass exclusion bound strengthens as a coupling increases. Naturally, one can ask about the maximum mass bound for a perturbative new coupling. Table.~\ref{tab:PertLim} shows the maximum (extrapolated) limits for all possible perturbative couplings, assuming only one nonzero coupling (see Ref.~\cite{Allwicher:2021rtd} for perturbative unitarity bounds on Yukawa couplings). Of course, the limits weaken if more than one coupling is nonzero and BR$($LQ $\to\ell j)$ is reduced.

\section{Summary and Discussions}\label{sec:conclu}
\noindent
Since LQs play essential roles in a wide range of BSM scenarios, it is crucial to interpret the experimental limits on them properly. In this paper, we revisited the exclusion limits on sLQs from the latest LHC $\mu \mu jj$ data. The pair-production searches at the LHC usually obtain largely model-agnostic exclusion bounds on sLQ masses. On the other hand, the high-$p_{\rm T}$-tails of the dilepton (or monolepton $+~\slashed E_{\rm T}$) resonance search data provide bounds on the sLQ-quark-lepton Yukawa couplings. Our study showed that, in some cases,  the LHC data exclude more parameter space than what we get by putting the above limits together. 

Our study essentially expanded the scope of Ref.~\cite{Mandal:2015vfa}, which argued for a systematic combination of events from the signal processes that could produce the same (or experimentally indistinguishable) final states. There, it was shown how one could include single-production events in the pair production signal (and vice versa) to obtain bounds on the Yukawa couplings from the pair production (dilepton-dijet) data. In our analysis, we extended the scope of the signal and included events from pair and single productions, $t$-channel sLQ exchange and its interference with the SM background processes. Our study is comprehensive as we considered all sLQs that can couple with a muon and a quark. Our results point to some interesting observations. First, the $\m\m jj$-recast limits are very restrictive and, in some cases, are the strongest. Second, when calculating the model-independent limits (i.e., assuming the $y\to 0$ limit), one ignores the QED-mediated contributions. However, Table~\ref{tab:SLQYukabcd} showed that the QED effects were not negligible in all cases, especially when sLQs carry higher electric charges. Moreover, the model-independent limits on the doublet and triplet sLQs are stronger than that on the singlet $S_1$ that decays to the $\m j$ final state with $100\%$ BR. This happens because there are multiple components of the doublet and triplet sLQs which contribute to the same final state and hence contribute to the limits.

\section*{Acknowledgements}\label{sec:aknow}
\noindent
We thank Anirudhan A. Madathil for the discussions on muon $g-2$. We also thank Gokul and Rithika for helping with \textsc{FeynRules} model files. R.S. acknowledges the Prime Minister's Research Fellowship (PMRF ID: 0802000).

\appendix

\section{Notation for the Yukawa couplings}\label{appendix:notation}
\noindent
We use an intuitive notation for the LQ Yukawa couplings in our paper and the \textsc{FeynRules} model files, publicly available at \href{https://github.com/rsrchtsm/LQ_Models}{GitHub}. In this notation, the sLQ-quark-lepton Yukawa couplings are generically denoted as $y_{pq,\,ij}^{AB}$, where the subscript $p$ indicates the $SU(2)$ representation of the sLQ; $q$ is $1$ or $2$ depending on whether there is a tilde ($\sim$) or a bar ($-$) over the LQ, it is zero otherwise; $i$ and $j$ denote the quark and the lepton generations, respectively. The superscripts $A$ and $B$ show the quark and lepton chiralities, respectively. For example, $y^{RR}_{11,\,22}$ indicates the coupling of $\widetilde S_1$ with a right-handed second-generation quark and a right-handed second-generation lepton. Similarly, $y^{RL}_{21,\,12}$ is the coupling of the $SU(2)$-doublet $\widetilde R_2$ with a right-handed first-generation quark and left-handed second-generation lepton.

\def\bibfont{\small}
\bibliography{References}{}
\bibliographystyle{JHEPCust}

\end{document}